\documentclass[12pt]{article}
\usepackage{epsfig}
\usepackage{amssymb}
\usepackage{amsmath}
\usepackage{epstopdf}
\usepackage{graphicx} 
\usepackage{color} 
\usepackage[caption=false]{subfig}
\usepackage{bm}  
\begin{document}
\thispagestyle{empty}
\begin{center}
\Large{{\bf Few-body techniques using momentum space for bound and continuum states}} \\
\vspace{1cm}
\Large{{\bf Marcelo Takeshi Yamashita}} \\
\vspace{0.6cm}
{Institute of Theoretical Physics - S\~ao Paulo State University}
\newpage
\large{{\bf Preface}}
\end{center}
\noindent
These notes were written for a set of three lectures given in a school at the Max Planck Institute for the Physics of Complex Systems in October/2017 before the workshop ``Critical Stability of Quantum Few-Body Systems''. These lectures are primarily dedicated to the students and represent a very idiosyncratic vision of the author, mainly in the last part of the text related to applications. These notes are only a tentative to show a technique, among many others, to solve problems in a very rich area of the contemporary physics - the Few-Body Physics - many times unknown by a considerable part of the students.
\vfill
\begin{center}
\large{October/2017}
\end{center}
\newpage

\tableofcontents

\section{Introduction}

The first question we probably think by reading ``Few-Body Physics'' along the title \footnote{A curiosity to mention here is that we usually hear that the classical few-body problem doesn't have a solution for three or more particles, as demonstrated by Henri Poincar\'e when he was 35 years old. In fact, the $N$-body classical problem was already solved in 1991 by Qiudong Wang, at that time, a student in the beginning of his PhD \cite{QDW}.} is what is the meaning of the word ``Few-Body''. This term may seem a little vague and use it in order to define an area of the physics also seems incautious, mainly by the use of a so subjective word like ``few''. In our case, the word ``body'' corresponds to any particle that may be present in several contexts of the physics like, e.g., quarks, protons, neutrons, atoms or molecules. Each of these different constituents should have a common characteristic: they should be a very defined object. They should not be treated as an approximation like in a mean-field theory, for example. Thus the word ``few'' should be understood taking into account the technical and computational difficulties that appear as the number of particles increase. A few-body system may represent 3, 4, 5, etc. particles since the individuality of each object is respected.

These lecture notes, far from being definitive, are written as follows: initially, we present a short description of the scattering theory and also introduce the Faddeev equations. This section may be easily found in textbooks, but it will serve to define the notations used along the text. Then, in the next section, we will describe how to calculate three-body bound and scattering states. The last sections will be used to show some applications of the formalism developed here.

\section{A very brief introduction to scattering theory}

The main objective of this section is to fix some concepts and notations related to the formal scattering theory that will be useful in the other sections. For a complete view of this subject we recommend the books of C. J. Joachain \cite{joachain}, R. G. Newton \cite{newton} and \cite{ziegelmann}. Here, we will follow closely the reference \cite{joachain}.

\subsection{General scattering formalism}

Heisenberg and Schroedinger pictures are two equivalent ways to treat the dynamics of a quantum system. In the Heisenberg picture the dynamics is given by the time evolution of the operators that represent the observables of the system and, in this case, the state vectors do not depend on time. On the other hand, in the Schroedinger picture the operators do not depend on time and the time evolution is given by the state vectors. There is, however, an intermediate description called interaction picture, where the dynamics is mixed between operators and state vectors. We will use here, for convenience, the interaction picture to study the scattering process.

Consider the time-independent Hamiltonian, $H$, and assume that it can be written as a sum of two terms $H=H_0+V$, where $V$ is a time independent local potential and $H_0$ is the free Hamiltonian satisfying $H_0\Phi=E\Phi$. The idea of this intermediate description is to separate the free part, associated uniquely to $H_0$, from the complete movement of the system. Consider, for instance, the unitary evolution operator $U_0$, defined as
\begin{equation}
U_0(t,t_0)\equiv e^{-\frac{i}{\hbar}H_0(t-t_0)},
\end{equation}
and the state vector $\Psi(t)\equiv U(t,t_0)\Psi(t_0)$, where $U(t,t_0)\equiv e^{-\frac{i}{\hbar}H(t-t_0)}$. Let us apply $U_0$ to $\Psi(t)$ and define:
\begin{eqnarray}
\Psi^{({\rm I})}(t)\equiv U_0^\dagger(t,t_0)\Psi(t)=U_0^\dagger(t,t_0)U(t,t_0)\Psi(t_0),
\label{psiidef}
\end{eqnarray}
where the superscripts mean interaction. Differentiating $\Psi^{({\rm I})}(t)$ with respect to the time, we have
\begin{eqnarray}
i\hbar\frac{\partial\Psi^{({\rm I})}(t)}{\partial t}=V(t)\Psi^{({\rm I})}(t),
\label{tomonaga}
\end{eqnarray}
where 
\begin{equation}
V(t)\equiv U_0^\dagger(t,t_0)VU_0(t,t_0), 
\label{vt}
\end{equation}
where it was used that the operator $U_0$ is unitary. Eq. (\ref{tomonaga}) is known as Tomonaga-Schwinger equation\cite{tomonaga,schwinger}. 

\noindent\rule[0.5ex]{\linewidth}{1pt}
{\bf Exercise 1:} Verify Eq. (\ref{tomonaga}).

\noindent\rule[0.5ex]{\linewidth}{1pt}

Comparing with Eq. (\ref{vt}), note that the time evolution of the system is mixed between the state vectors, as in the Schroedinger picture, and the operators, as in the Heisenberg picture. It is worth to note that the evolution of the operators given by
\begin{equation}
i\hbar\frac{dA^{({\rm I})}(t)}{dt}=[A^{({\rm I})}(t),H_0]
\end{equation}
depends only on $H_0$. The evolution of the state vectors $\Psi^{({\rm I})}$ (conveniently modified to include the time evolution corresponding to $H_0$), Eq. (\ref{tomonaga}), depends only on $V$. Based on Schwinger-Tomonaga equation we can define a time evolution operator in the interaction picture as
\begin{equation}
\Psi^{({\rm I})}(t)\equiv U_{({\rm I})}(t,t^\prime)\Psi^{({\rm I})}(t^\prime).
\label{ui}
\end{equation}
We may, then, write Eq. (\ref{tomonaga}) as
\begin{eqnarray}
i\hbar\frac{\partial}{\partial t} U_{({\rm I})}(t,t^\prime)\Psi^{({\rm I})}(t^\prime)=V(t)U_{({\rm I})}(t,t^\prime)\Psi^{({\rm I})}(t^\prime).
\label{eq7}
\end{eqnarray}
As Eq. (\ref{eq7}) is valid for any $t^\prime$, we may deduce that:
\begin{equation}
i\hbar\frac{\partial U_{({\rm I})}(t,t^\prime)}{\partial t}=V(t)U_{({\rm I})}(t,t^\prime),
\end{equation}
which can be written as a Volterra-type integral equation
\begin{equation}
U_{({\rm I})}(t,t^\prime)=I-i\hbar\int_{t^\prime}^tV(t^{\prime\prime})U_{({\rm I})}(t,t^{\prime\prime})dt^{\prime\prime},
\end{equation}
where $I$ is the identity operator. From the definition given by Eq. (\ref{psiidef}) we can write
\begin{equation}
\Psi^{({\rm I})}(t)=U_0^\dagger(t,t_0)\Psi(t)=U_0^\dagger(t,t_0)U(t,t^\prime)\Psi(t^\prime),
\end{equation}
from Eq. (\ref{psiidef}) and using the fact that the evolution operator is unitary, we have
\begin{equation}
\Psi^{({\rm I})}(t)=U_0^\dagger(t,t_0)U(t,t^\prime)U_0(t^\prime,t_0)U_0^\dagger(t^\prime,t_0)\Psi(t^\prime)=U_0^\dagger(t,t_0)U(t,t^\prime)U_0(t^\prime,t_0)\Psi^{({\rm I})}(t^\prime),
\end{equation}
which comparing with Eq. (\ref{ui}) gives
\begin{equation}
U_{({\rm I})}(t,t^\prime)=U_0^\dagger(t,t_0)U(t,t^\prime)U_0(t^\prime,t_0).
\label{ui2}
\end{equation}

In a scattering problem, usually we have a particle beam that is prepared in a remote past, interact with a target and finally reaches detectors positioned in an asymptotic reagion in a far future. Then, we are indeed interested in the limits where $t,t^\prime\rightarrow\pm\infty$. These limits, however, are not well defined as depicted by eq. (\ref{ui2}). One of the ways to calculate these limits is to use the recipe developed by Gell-Mann and Goldberger \cite{gellmann}. The method consists essentially in eliminate the oscillatory behaviour of the operator multiplying it by an exponential function. The limit to be calculated may be written as
\begin{eqnarray}
&&\lim_{t\to-\infty}F(t)=\lim_{\epsilon\to0^+}\epsilon\int_{-\infty}^0 e^{\epsilon t^\prime}F(t^\prime)dt^\prime, \label{gell1} \\
&&\lim_{t\to+\infty}F(t)=\lim_{\epsilon\to0^+}\epsilon\int_0^\infty e^{-\epsilon t^\prime}F(t^\prime)dt^\prime. \label{gell2}
\end{eqnarray}

Eqs. (\ref{gell1}) and (\ref{gell2}) may be used to calculate the limits of the evolution operators for $t,t^\prime\rightarrow\pm\infty$. Now, we can define the so called M\o ller operators:
\begin{eqnarray}
\Omega^{(\pm)}\equiv U_{({\rm I})}(0,\mp\infty) \label{omega}\\
\Omega^{(\pm)\dagger}\equiv U_{({\rm I})}(\mp\infty,0). \label{omegad}
\end{eqnarray}

If we denote by $\Phi_i$, where $i=\alpha,\beta...$ represents a set of quantum numbers, we see that $\Phi_i$ is an eigenvalue of the operator $\Omega^{(\pm)}$, i.e. this operator converts an eigenstate of the free Hamiltonian, $H_0$, into an eigenstate $\Psi_i(t=0)$ of the full Hamiltonian
\begin{equation}
|\Psi_\alpha^{(\pm)}\rangle=\Omega^{(\pm)}|\Phi_\alpha\rangle, \,\,\, \langle\Psi_\alpha^{(\pm)}|=\langle\Phi_\alpha|\Omega^{(\pm)\dagger}.
\label{phipsi}
\end{equation}
As $\langle\Phi_\beta|\Phi_\alpha\rangle=\delta_{\beta\alpha}$, we may write
\begin{equation}
\Omega^{(\pm)}=\sum_{j=\alpha,\beta,\dots}|\Psi_j^{(\pm)}\rangle\langle\Phi_j|.
\end{equation}

Returning to Eqs. (\ref{ui2}), (\ref{gell1}) and (\ref{gell2}) with $t_0=0$ we may write the M\o ller operators as
\begin{eqnarray}
\Omega^{(\pm)}&=&\lim_{t\to\mp\infty}U_{({\rm I})}(0,t)=\lim_{t\to\mp\infty}U(0,t)U_0(t,0)\\
&=&\lim_{\epsilon\to0^+}\mp\epsilon\int_0^{\mp\infty} e^{\pm\epsilon t}e^{\frac{i}{\hbar}Ht}e^{-\frac{i}{\hbar}H_0t}dt.
\label{mollerint}
\end{eqnarray}

We will now assume that the states $\Phi_\alpha$ form a full basis\footnote{This assumption is due by considering several conditions that were not explicitly discussed here, e.g.: we are assuming that the Hamiltonian of the system doesn't have bound states and that exists only one scattering channel. These assumptions, besides very restrictive, don't interfere on the understanding of the physical problem we are considering. Check Ref. \cite{joachain,newton} for a detailed and complete description.}, i.e.
\begin{equation}
\sum_{j=\alpha,\beta,\dots}|\Phi_j\rangle\langle\Phi_j|=I.
\label{comp}
\end{equation} 
Inserting Eq. (\ref{comp}) into Eq. (\ref{mollerint}), we have (\underline{we will simplify the notation assuming $\hbar=1$})
\begin{eqnarray}
\nonumber
\Omega^{(\pm)}&=&\lim_{\epsilon\to0^+}\mp\epsilon\sum_{j=\alpha,\beta,\dots}\int_0^{\mp\infty} e^{\pm\epsilon t}e^{iHt}|\Phi_j\rangle\langle\Phi_j|e^{-iH_0t}dt.\\
&=&\lim_{\epsilon\to0^+}\mp\epsilon\sum_{j=\alpha,\beta,\dots}\int_0^{\mp\infty} e^{\pm\epsilon t}e^{iHt}|\Phi_j\rangle\langle\Phi_j|e^{-iE_j t}dt.
\label{mollerint1}
\end{eqnarray}
This integral can be easily calculated giving
\begin{equation}
\Omega^{(\pm)}=\lim_{\epsilon\to0^+}\sum_{j=\alpha,\beta,\dots}\frac{\pm i\epsilon}{E_j-H\pm i\epsilon}|\Phi_j\rangle\langle\Phi_j|,
\label{mollerint2}
\end{equation}
replacing Eq. (\ref{mollerint2}) in (\ref{phipsi}) gives
\begin{equation}
|\Psi_\alpha^{(\pm)}\rangle=\lim_{\epsilon\to0^+}\frac{\pm i\epsilon}{E_\alpha-H\pm i\epsilon}|\Phi_\alpha\rangle,
\label{hom}
\end{equation}
where it was used the orthonormality of $\Phi$. We may still write this equation as
\begin{eqnarray}
\nonumber
&&|\Psi_\alpha^{(\pm)}\rangle=\lim_{\epsilon\to0^+}\frac{1}{E_\alpha-H\pm i\epsilon}(E_\alpha-H_0-V+V\pm i\epsilon)|\Phi_\alpha\rangle\\
&&|\Psi_\alpha^{(\pm)}\rangle=|\Phi_\alpha\rangle+\lim_{\epsilon\to0^+}\frac{V}{E_\alpha-H\pm i\epsilon}|\Phi_\alpha\rangle.
\label{ls1}
\end{eqnarray}
The Green operator, $G^{(\pm)}$, is defined as
\begin{equation}
G^{(\pm)}\equiv\lim_{\epsilon\to0^+}\frac{1}{E_\alpha-H\pm i\epsilon}.
\end{equation}
It is called resolvent of the operator $H$ the function given by $G(z)=\frac{1}{z-H}$ (the free Green operator, $G_0(z)$, is obtained replacing $H\to H_0$). The free Green function, can be obtained, for example, in the configuration space $\{\vec{r}\}$ calculating the matrix element
\begin{equation}
\langle\vec{r}|\frac{1}{E_\alpha-H_0\pm i\epsilon}|\vec{r}\,^\prime\rangle=\int d\vec{k}^\prime\frac{1}{E_\alpha-E_{k^\prime}\pm i\epsilon}\langle\vec{r}|\Phi_{k^\prime}\rangle\langle\Phi_{k^\prime}|\vec{r}\,^\prime\rangle,
\end{equation}
making the following substitutions $\vec{r}-\vec{r}\,^\prime\equiv\vec{R}$,  $E_\alpha\equiv\frac{k^2}{2m}$, $E_{k^\prime}\equiv\frac{k^{\prime2}}{2m}$ and $\langle\vec{r}|\Phi_{k^\prime}\rangle\equiv\frac{1}{(2\pi)^{3/2}}e^{i\vec{k}^\prime\cdot\vec{r}}$ we have that the matrix element, after performing the angular integration, may be written as
\begin{eqnarray}
\nonumber
&&\langle\vec{r}|\frac{1}{E_\alpha-H_0\pm i\epsilon}|\vec{r}\,^\prime\rangle=\\
&&-2m\frac{1}{16\pi^2iR}\int_{-\infty}^{\infty}dk^\prime\left\{e^{ik^\prime R}\left[\frac{1}{k^\prime+k}+\frac{1}{k^\prime-k}\right]-e^{-ik^\prime R}\left[\frac{1}{k^\prime+k}+\frac{1}{k^\prime-k}\right]\right\},
\end{eqnarray}
where we used the fact that the integrand is even to extend the integration from $-\infty$ to $\infty$. This integral may be solved using the Cauchy's integral formula. Then, we can define
\begin{eqnarray}
&&I_1\equiv\oint_{C_1}dk^\prime e^{ik^\prime R}\left[\frac{1}{k^\prime+k}+\frac{1}{k^\prime-k}\right],\\
&&I_2\equiv\oint_{C_2}dk^\prime e^{-ik^\prime R}\left[\frac{1}{k^\prime+k}+\frac{1}{k^\prime-k}\right].
\end{eqnarray}
The integral $I_1$ should have its contour $C_1$ closed in the anticlockwise direction in the upper-half plane and $C_2$ should be closed in the clockwise direction in the bottom-half plane.  

\pagebreak
\noindent\rule[0.5ex]{\linewidth}{1pt}
{\bf Exercise 2:} Justify why the contours $C_1$ and $C_2$ should be closed as stated above.

\noindent\rule[0.5ex]{\linewidth}{1pt}

We can, however, make $C_1$ or $C_2$ that contain, or not, the poles of the denominator, where each situation represents a different result to the integrals as showed in table \ref{cauchy}.
\begin{table}[htb!]
\centering 
\begin{tabular}{c|c|c|c|c} 
\hline\hline 
 & $C_1$ & $C_2$ & $I_1$ & $I_2$ \\ [0.5ex] 
\hline
1 & poles $\pm k$     & 0                    & $2\pi i\left(e^{ikR}+e^{-ikR}\right)$ & 0 \\
2 & pole $+k$           & pole $-k$      & $2\pi ie^{ikR}$ & $-2\pi ie^{ikR}$ \\
3 & pole $-k$            & pole $+k$     & $2\pi ie^{-ikR}$ & $-2\pi ie^{-ikR}$ \\
4 & 0                          & poles $\pm k$    & 0 & -$2\pi i\left(e^{ikR}+e^{-ikR}\right)$ \\ [1ex] 
\hline 
\end{tabular}
\caption{Results for the integrals $I_1$ and $I_2$ according to the inclusion, or not, of the poles $\pm k$ inside the contours $C_1$ and $C_2$ closed by a semicircle, respectively, in the anti and clockwise directions.}
\label{cauchy}
\end{table}

In Table \ref{cauchy} we see that the combination 2 is the only one that gives for large $R$ an outgoing spherical wave. Thus, the matrix element we are calculating is given by
\begin{eqnarray}
\nonumber
\langle\vec{r}|\frac{1}{E_\alpha-H_0\pm i\epsilon}|\vec{r}\,^\prime\rangle&=&-\frac{2m}{\hbar^2}\frac{1}{16\pi^2iR}\left(2\pi ie^{ikR}+2\pi ie^{ikR}\right)=\\
&=&\frac{2m}{\hbar^2}\left(-\frac{1}{4\pi}\frac{e^{ik|\vec{r}-\vec{r}\,^\prime|}}{|\vec{r}-\vec{r}\,^\prime|}\right)\equiv\frac{2m}{\hbar^2}G_0^{(+)}(\vec{r},\vec{r}\,^\prime),
\end{eqnarray}
where
\begin{equation}
G_0^{(+)}(\vec{r},\vec{r}\,^\prime)\equiv-\frac{1}{4\pi}\frac{e^{ik|\vec{r}-\vec{r}\,^\prime|}}{|\vec{r}-\vec{r}\,^\prime|}
\label{green}
\end{equation}
is the free Green function for an outgoing spherical wave. For an incoming spherical wave we write
\begin{equation}
G_0^{(-)}(\vec{r},\vec{r}\,^\prime)\equiv-\frac{1}{4\pi}\frac{e^{-ik|\vec{r}-\vec{r}\,^\prime|}}{|\vec{r}-\vec{r}\,^\prime|}.
\end{equation}

After having defined the Green operator and the Green function we want to derive the Lippmann-Schwinger equation. Let us start using the identity $\frac1A-\frac1B=\frac1B(B-A)\frac1A$ to write
\begin{equation}
G^{(\pm)}=G_0^{(\pm)}+G_0^{(\pm)}VG^{(\pm)}. 
\label{gg0}
\end{equation}

\noindent\rule[0.5ex]{\linewidth}{1pt}
{\bf Exercise 3:} Verify Eq. (\ref{gg0}).

\noindent\rule[0.5ex]{\linewidth}{1pt}

Replacing this identity in Eq. (\ref{ls1}), we have
\begin{eqnarray}
\nonumber
|\Psi_\alpha^{(\pm)}\rangle&=&|\Phi_\alpha\rangle+G^{(\pm)}V|\Phi_\alpha\rangle
=|\Phi_\alpha\rangle+\left(G_0^{(\pm)}+G_0^{(\pm)}VG^{(\pm)}\right)V|\Phi_\alpha\rangle\\
&=&|\Phi_\alpha\rangle+G_0^{(\pm)}V\left(|\Phi_\alpha\rangle+G^{(\pm)}V|\Phi_\alpha\rangle\right),
\label{als}
\end{eqnarray}
where the term that appears between the parenthesis after the last equality is $|\Psi_\alpha^{(\pm)}\rangle$. Then, we may write
\begin{equation}
|\Psi_\alpha^{(\pm)}\rangle=|\Phi_\alpha\rangle+G_0^{(\pm)}V|\Psi_\alpha^{(\pm)}\rangle,
\label{LS}
\end{equation}
which is the famous Lippmann-Schwinger equation.

\noindent\rule[0.5ex]{\linewidth}{1pt}
{\bf Exercise 4:} Starting from the Lippmann-Schwinger equation, show that the asymptotic form of the scattering wave functions $\Psi_\alpha^{(\pm)}(\vec{r})$ by a local potential $V(\vec{r})$ reduces, for $k\to0$, to
\begin{equation}
\nonumber
\Psi_\alpha^{(\pm)}(\vec{r})\to1-\frac{a}{r},
\end{equation}
where the scattering length is given by
\begin{equation}
\nonumber
a=\frac{m}{2\pi\hbar^2}\int d^3rV(\vec{r})\Psi_\alpha^{(\pm)}(\vec{r}).
\end{equation}
Show that for a purely repulsive potential, $a>0$.

\noindent\rule[0.5ex]{\linewidth}{1pt}

Let us return to the M\o ller operators in order to define the operator and collision matrix that relate the state vector of the system in a remote past ($t\to-\infty$) to a distant future ($t\to\infty$). The collision operator is called $S$ and is defined as
\begin{equation}
S\equiv U_{({\rm I})}(+\infty,0)U_{({\rm I})}(0,-\infty)=\Omega^{(-)\dagger}\Omega^{(+)}.
\end{equation}
Calculating the matrix element of the operator $S$ between two states $\Phi_\alpha$ and $\Phi_\beta$ we have that
\begin{equation}
\langle\Phi_\beta|S|\Phi_\alpha\rangle=\langle|\Phi_\beta|\Omega^{(-)\dagger}\Omega^{(+)}|\Phi_\alpha\rangle=\langle\Psi_\beta^{(-)}|\Psi_\alpha^{(+)}\rangle.
\label{smatrix}
\end{equation}
Adding and subtracting $\langle\Psi_\beta^{(+)}|\Psi_\alpha^{(+)}\rangle$, Eq. (\ref{smatrix}) may still be written as
\begin{eqnarray}
\nonumber
\langle\Phi_\beta|S|\Phi_\alpha\rangle&=&\langle\Psi_\beta^{(+)}|\Psi_\alpha^{(+)}\rangle+\langle\Psi_\beta^{(-)}-\Psi_\beta^{(+)}|\Psi_\alpha^{(+)}\rangle\\
&=&\delta_{\alpha\beta}+\langle\Psi_\beta^{(-)}-\Psi_\beta^{(+)}|\Psi_\alpha^{(+)}\rangle.
\label{smatrix1}
\end{eqnarray}

Using the hermiticity of $H$ and $V$ we may write Eq. (\ref{ls1}) as $\langle\Psi_\beta^{(\pm)}|=\langle\Phi_\beta|+\lim_{\epsilon\to0^+}\langle\Phi_\beta|\frac{V}{E_\beta-H\mp i\epsilon}$ and replacing it in Eq. (\ref{smatrix1}), we have
\begin{eqnarray}
\nonumber
\langle\Phi_\beta|S|\Phi_\alpha\rangle&=&\delta_{\alpha\beta}+\lim_{\epsilon\to0^+}\langle\Phi_\beta|\frac{V}{E_\alpha-H+ i\epsilon}-\frac{V}{E_\beta-H- i\epsilon}|\Psi_\alpha^{(+)}\rangle\\ \nonumber
&=&\delta_{\alpha\beta}+\lim_{\epsilon\to0^+}\left(\frac{1}{E_\beta-E_\alpha+ i\epsilon}-\frac{1}{E_\beta-E_\alpha- i\epsilon}\right)\langle\Phi_\beta|V|\Psi_\alpha^{(+)}\rangle\\ 
&=&\delta_{\alpha\beta}-\lim_{\epsilon\to0^+}\frac{2i\epsilon}{(E_\beta-E_\alpha)^2+ \epsilon^2}\langle\Phi_\beta|V|\Psi_\alpha^{(+)}\rangle.
\label{smatrix2}
\end{eqnarray}

Note that one of the ways to represent the Dirac delta function is from the limit given by:
\begin{equation}
\delta(x-x_0)=\frac1\pi\lim_{\epsilon\to0^+}\frac{\epsilon}{(x-x_0)^2+ \epsilon^2}.
\end{equation}
This function returns 0 for every $x\neq x_0$ and infinity for $x=x_0$. The integral of this function over all space should be equal to 1 -  for this reason the factor $\frac1\pi$ is multiplying the function. Then, we may replace the limit which appears in Eq. (\ref{smatrix2}) giving
\begin{equation}
\langle\Phi_\beta|S|\Phi_\alpha\rangle=\delta_{\alpha\beta}-2\pi i\delta(E_\beta-E_\alpha)\langle\Phi_\beta|V|\Psi_\alpha^{(+)}\rangle.
\label{smatrix3}
\end{equation}

The transition operator $T$ may be defined as
\begin{equation}
T\equiv V+VG^{(\pm)}V.
\label{tmatrixg}
\end{equation}

Replacing $|\Psi_\alpha^{(\pm)}\rangle=|\Phi_\alpha\rangle+G^{(\pm)}V|\Phi_\alpha\rangle$ (the first equality of Eq. (\ref{als})) in Eq. (\ref{smatrix3}), we can relate the $T$-matrix elements, $\langle\Phi_\beta|T|\Phi_\alpha\rangle$, to the $S$ matrix as
\begin{equation}
\langle\Phi_\beta|S|\Phi_\alpha\rangle=\delta_{\alpha\beta}-2\pi i\delta(E_\beta-E_\alpha)\langle\Phi_\beta|T|\Phi_\alpha\rangle.
\end{equation}

The transition operator may also be written as
\begin{eqnarray}
T=V+VG_0^{(\pm)}T=V+TG_0^{(\pm)}V.
\label{tmatrixg0}
\end{eqnarray}
Note that the equivalence of Eqs. (\ref{tmatrixg0}) and (\ref{tmatrixg}) is easily shown by iteration
\begin{eqnarray}
\nonumber
T&=&V+VG_0^{(\pm)}\left(V+VG_0^{(\pm)}\left(V+VG_0^{(\pm)}...\right)\right)\\
&=&V+V\left(G_0^{(\pm)}+G_0^{(\pm)}VG_0^{(\pm)}+...\right)V\\
&=&V+VG^{(\pm)}V,
\end{eqnarray}
where it was used Eq. (\ref{gg0}). Obviously, the second equality of Eq. (\ref{tmatrixg0}) may be demonstrated in the same manner.

\subsection{A pinch of integral equation theory (from the book of M. Masujima, Applied Mathematical Methods in Theoretical Physics \cite{masujima})}

This subsection describes very superficially a specific topic in the integral-equation theory. The aim of this short section is to show that the Lippmann-Schwinger equation for a two-body problem is a ``good'' equation with a unique solution. This same equation has some problems when applied to three particles. For a complete view of integral equation theory you may check Ref. \cite{masujima}. 

The matrix elements of the transition operator calculated in momentum space reads
\begin{eqnarray}
&&\langle\vec{p}|T|\vec{p}\,^\prime\rangle=\langle\vec{p}|V|\vec{p}\,^\prime\rangle+\int d^3p^{\prime\prime}\langle\vec{p}\,|V|\vec{p}\,^{\prime\prime}\rangle\frac{1}{E-\frac{p^{\prime\prime2}}{2m}\pm i\epsilon}\langle\vec{p}\,^{\prime\prime}|T|\vec{p}\,^\prime\rangle.
\label{Tintegral}
\end{eqnarray}
Successive iterations of the above equation returns the Born series for the $T$-operator. Eq. (\ref{Tintegral}) is classified as a Fredholm integral equation of second type, written generically as

\begin{equation}
\phi(x)=f(x)+\lambda\int_0^h dx^\prime K(x,x^\prime)\phi(x^\prime)\;\;\;(0\leq x\leq h).
\label{int}
\end{equation}

Assuming that, both $f$ and $K$ are square integrable which means
\begin{equation}
\|f\|^2<\infty\;\;{\rm and}\;\;\|K\|^2\equiv\int_0^hdx\int_0^hdx^\prime |K(x,x^\prime)|^2<\infty,
\label{si}
\end{equation}
let us try an iterative solution in $\lambda$ for this problem:
\begin{equation}
\phi(x)=\phi_0(x)+\lambda\phi_1(x)+\lambda^2\phi_2(x)+\cdots+\lambda^n\phi_n(x)+\cdots
\label{serie}
\end{equation}
Replacing Eq.(\ref{serie}) in (\ref{int}) we have that
\begin{eqnarray}
\phi_0(x)&=&f(x)\\
\phi_1(x)&=&\int_0^hdy_1K(x,y_1)\phi_0(y_1)=\int_0^hdy_1K(x,y_1)f(y_1)\\
\phi_2(x)&=&\int_0^hdy_2K(x,y_2)\phi_1(y_2)=\int_0^hdy_2\int_0^hdy_1K(x,y_2)K(y_2,y_1)f(y_1)\\
&\vdots&
\end{eqnarray}
So, in general we have
\begin{eqnarray}
\nonumber
\phi_n(x)&=&\int_0^hdy_nK(x,y_n)\phi_{n-1}(y_n)\\
&=&\int_0^hdy_n\int_0^hdy_{n-1}\cdots\int_0^hdy_1 K(x,y_n)K(y_n,y_{n-1})
\cdots K(y_2,y_1)f(y_1).
\end{eqnarray}

In order to set the boundary conditions let us first define:

\begin{equation}
A(x)\equiv \int_0^hdy|K(x,y)|^2\;\;{\rm thus}\;\;\|K\|^2=\int_0^hdxA(x).
\end{equation}

Then, using the Schwarz inequality, $|\langle u,v\rangle|^2\leq\|u\|\|v\|$, in the iteration serie $\phi_i\;\;(i=1,2,\dots,n,\dots)$ we have
\begin{eqnarray}
&&|\phi_1(x)|^2\leq A(x)\|f\|^2\\
&&|\phi_2(x)|^2\leq A(x)\|\phi_1\|^2\leq A(x)\|f\|^2\|K\|^2\\ \nonumber
\vdots\\
&&|\phi_n(x)|^2\leq A(x)\|f\|^2\|K\|^{2(n-1)},
\end{eqnarray}
thus 
\begin{equation}
|\phi_n(x)|\leq \sqrt{A(x)}\|f\|\|K\|^{(n-1)}\;\;\;(n=1,2,3,\dots)
\label{itserie}
\end{equation}

Now, returning to the iteration serie given by Eq. (\ref{serie}), $\phi(x)-f(x)=\lambda\phi_1(x)+\lambda^2\phi_2(x)+\cdots+\lambda^n\phi_n(x)+\cdots$, and using the triangle inequality ($\parallel\vec{u}+\vec{v}\parallel\leq\parallel\vec{u}\parallel+\parallel\vec{v}\parallel$) we may write:

\begin{eqnarray}
\nonumber
&&|\phi(x)-f(x)|=|\lambda\phi_1(x)+\lambda^2\phi_2(x)+\cdots|\leq |\lambda||\phi_1(x)|+|\lambda^2||\phi_2(x)|+\cdots=\\ \nonumber
&&=\sum_{n=1}^\infty |\lambda|^n|\phi_n(x)|\leq\sum_{n=1}^\infty |\lambda|^n|\sqrt{A(x)}\|f\|\|K\|^{(n-1)}=|\lambda|\sqrt{A(x)}\|f\|\sum_{n=0}^\infty|\lambda|^n\|K\|^n,
\end{eqnarray}
where in the last inequality we replaced Eq. (\ref{itserie}) and used the triangle inequality again. So, this serie converges if $|\lambda|\|K\|<1$, giving $\frac{1}{1-|\lambda|\|K\|}$.

We can also check the uniqueness of the solution. For this intent, we will see whether the homogeneous part of the integral equation has another solution than the trivial one. A non-trivial solution indicates that any multiple of this solution is also a solution, which means that we would have a myriad of solutions of the inhomogeneous equation. Let us consider the homogeneous problem:

\begin{equation}
\phi_H(x)=\lambda\int_0^hdyK(x,y)\phi_H(y).
\end{equation}
The Schwarz inequality ($|\langle\vec{u},\vec{v}\rangle|^2\leq|\langle\vec{u},\vec{u}\rangle|\langle\vec{v},\vec{v}\rangle$) reads:
\begin{equation}
|\phi_H(x)|^2\leq|\lambda|^2A(x)\|\phi_H\|^2.
\label{sch}
\end{equation}
The norm $\|\phi_H\|$ is explicitly written:
\begin{equation}
\|\phi_H\|^2=\int_0^hdx|\phi_H(x)|^2\leq|\lambda|^2\|\phi_H\|^2\int_0^hdxA(x)=|\lambda|^2\|K\|^2\|\phi_H\|^2.
\end{equation}
Note that to write the above inequality we used Eq. (\ref{sch}). Thus, we have that:
\begin{equation}
\|\phi_H\|^2(1-|\lambda|^2\|K\|^2)\leq0.
\end{equation}
As $|\lambda|\|K\|<1$, this inequality may only be satisfied if $\|\phi_H\|=0$ or $\phi_H\equiv0$ - only the trivial solution. Thus, the solution is unique.

In conclusion, for an integral equation $\phi(x)=f(x)+\lambda\int_0^hK(x,y)\phi(y)dy$ with a kernel belonging to the Hilbert-Schmidt class, $K(x,y)\in \mathcal{L}^2$, the Fredholm method can be applied and the equation has a unique continuous solution given by (for a detailed development of this part, please check \cite{masujima})
\begin{equation}
\phi(x)=f(x)-\lambda\int_0^hH(x,y;\lambda)f(y)dy,
\end{equation}
where $H(x,y;\lambda)=D(x,y;\lambda)/D(\lambda)$ with
\begin{eqnarray}
&&D(\lambda)=\sum_{n=0}^\infty\frac{(-\lambda)^n}{n\!}D_n\\
&&D_n=\int_0^hdx_1\cdots\int_0^hdx_nK
\left(\begin{array}{ccc}
x_1, & \cdots & x_n \\
x_1, & \cdots & x_n \end{array}\right); \;\;\; D_0=1
\end{eqnarray}
and
\begin{eqnarray}
&&D(x,y;\lambda)=\sum_{n=0}^\infty\frac{(-\lambda)^n}{n\!}C_n(x,y)\\
&&C_n(x,y)=\int_0^hdx_1\cdots\int_0^hdx_nK
\left(\begin{array}{ccc}
x_1, & \cdots & x_n \\
x_1, & \cdots & x_n \end{array}\right); \;\;\; C_0(x,y)=-K(x,y),
\end{eqnarray}
where
\begin{eqnarray}
K
\left(\begin{array}{ccc}
x_1, & \cdots & x_n \\
x_1, & \cdots & x_n \end{array}\right)\equiv
\left|\begin{array}{cccc}
K(z_1,w_1) & K(z_1,w_2) & \cdots & K(z_1,w_n) \\
K(z_2,w_1) & K(z_2,w_2) & \cdots & K(z_2,w_n) \\ 
\cdot      &            &        &            \\
\cdot      &            &        &            \\
\cdot      &            &        &            \\
K(z_n,w_1) & K(z_n,w_2) & \cdots & K(z_n,w_n) 
\end{array}\right|
\end{eqnarray}

It is now interesting to return to the Lippmann-Schwinger equation (\ref{LS}). In configuration space it is written as:
\begin{equation}
\psi(\vec{r})=\phi(\vec{r})+\lambda\int K(\vec{r},\vec{r}\,^\prime)\psi(\vec{r}\,^\prime) d\vec{r}\,^\prime,
\end{equation}
with $K(\vec{r},\vec{r}\,^\prime)=G_0^{(+)}(\vec{r},\vec{r}\,^\prime)V(\vec{r}\,^\prime)$ 
\begin{equation}
K(\vec{r},\vec{r}\,^\prime)=-\frac{1}{4\pi}\frac{e^{ik|\vec{r}-\vec{r}\,^\prime|}}{|\vec{r}-\vec{r}\,^\prime|}V(\vec{r}\,^\prime),
\end{equation}
where we replaced $G_0^{(+)}(\vec{r},\vec{r}\,^\prime)$ given by Eq. (\ref{green}) (note that we inserted a $\lambda$ in front of the integral). In order to investigate if the kernel is square integrable we may consider $k={\rm Re}(k)+i{\rm Im}(k)$. Then, the kernel $K \in \mathcal{L}^2$ if
\begin{eqnarray}
\int d\vec{r}\int d\vec{r}\,^\prime|K(\vec{r},\vec{r}\,^\prime)|^2=\frac{1}{16\pi^2}\int d\vec{r}d\vec{r}\,^\prime \frac{e^{-2{\rm Im}(k)|\vec{r}-\vec{r}\,^\prime|}}{|\vec{r}-\vec{r}\,^\prime|^2}|V(\vec{r}\,^\prime)|^2<\infty,
\end{eqnarray}
substituting $\vec{s}=\vec{r}-\vec{r}\,^\prime$ we have
\begin{eqnarray}
\frac{1}{16\pi^2}\int d\vec{r}\,^\prime |V(\vec{r}\,^\prime)|^2\int d\vec{s}\frac{e^{-2{\rm Im}(k)s}}{s^2}<\infty,
\end{eqnarray}
if ${\rm Im}(k)>0$ and $\int d\vec{r}\,^\prime |V(\vec{r}\,^\prime)|^2<\infty$ the Fredholm method can be applied. 

However, consider a real $k$. Then, in this situation, we cannot apply the Fredholm method. This problem, which basically appears in the two-body problem, can be circumvented quite easily making the following substitutions and multiplying the Lippmann-Schwinger equation by $\sqrt{|V(\vec{r})|}$
\begin{eqnarray}
\nonumber
&&V(\vec{r})=|V(\vec{r})|\eta(\vec{r})\;\;(\eta(\vec{r})=1\;{\rm for}\;V(\vec{r})>0\;\;{\rm or}\;\;\eta(\vec{r})=-1\;{\rm for}\;V(\vec{r})<0),\\ \nonumber
&&\tilde{\psi}(\vec{r})=\sqrt{|V(\vec{r})|}\psi(\vec{r}),\\ \nonumber
&&\tilde{\phi}(\vec{r})=\sqrt{|V(\vec{r})|}\phi(\vec{r})
\end{eqnarray}
the kernel may be written as
\begin{equation}
\tilde{K}(\vec{r},\vec{r}\,^\prime)=-\frac{1}{4\pi}\sqrt{|V(\vec{r})|}\frac{e^{ik|\vec{r}-\vec{r}\,^\prime|}}{|\vec{r}-\vec{r}\,^\prime|}\sqrt{V(\vec{r}\,^\prime)}\eta(\vec{r}\,^\prime),
\label{ktilde}
\end{equation}
and the Lippmann-Schwinger equation is rewritten as
\begin{equation}
\tilde{\psi}(\vec{r})=\tilde{\phi}(\vec{r})+\lambda\int \tilde{K}(\vec{r},\vec{r}\,^\prime)\tilde{\psi}(\vec{r}\,^\prime) d\vec{r}\,^\prime.
\end{equation}

Now, the new kernel, $\tilde{K}$, given by Eq. (\ref{ktilde}) is square integrable in the limit of ${\rm Im}(k)=0$,
\begin{equation}
\frac{1}{16\pi^2}\int d\vec{r}d\vec{r}\,^\prime|V(\vec{r})||V(\vec{r}\,^\prime)|\frac{1}{|\vec{r}-\vec{r}\,^\prime|^2},
\end{equation}
if $\int d\vec{r}|V(\vec{r})|<\infty$ and $\int d\vec{r}\,^\prime\frac{|V(\vec{r}\,^\prime)|}{|\vec{r}-\vec{r}\,^\prime|^2}<\infty$.

The general mathematical discussion made until this point is important because when we are considering a two-body problem it is possible to separate the movement of the center of mass of the system defining:
\begin{eqnarray}
\vec{p}\equiv\frac{m_2\vec{p}_1+m_1\vec{p}_2}{m_1+m_2}\;\;\;\mu\equiv\frac{m_1m_2}{m_1+m_2},
\end{eqnarray}
in such a way the Hamiltonian of the system may be written as
\begin{equation}
H=\frac{p^2}{2\mu}+\frac{P^2}{2(m_1+m_2)}+V,
\end{equation}
where $p$ is the relative momentum between particles 1 and 2 (with masses $m_1$ and $m_2$), $\mu$ is the reduced mass and $P$ is the center of mass momentum. If the potential is translational invariant, the motion of the center of mass is just a free motion, then, we may just disregard it and concentrate only in the relative motion. As a consequence, we can reduce the degrees of freedom of the problem to only 3 and the equation we have to solve is analogous to the Lippmann-Schwinger equation written in this subsection. Thus, the solution is well behaved and unique. However, for a three-body problem we will see that this is not true and we will need to sophisticate a bit the theory. We will do this in the next subsection.

\subsection{The Faddeev equations}

What we showed in the last subsection is only an overview of the mathematics involved in the Lippmann-Schwinger equation. It is really far from the rigorous proof demanded by a mathematician. In this section, we show a problem involving the kernel of the Lippmann-Schwinger equation associated to the existence of more than one channel, which occurs when treating a three-body problem. Consider, for instance, a proton-deuteron scattering producing three free particles (two protons and a neutron). In this case, we might have infinite decay channels, for example: the proton produces an excited deuteron, which decays further and the proton emerges in the direction of the other proton interacting again. This is only one example and it is not difficult to imagine many others. 

As showed in the simple case of a two-body problem, it is possible to separate the movement of the center of mass reducing the degrees of freedom. This also happens for a three-body problem. Then, by using the Jacobi coordinates we can write the three-body Hamiltonian as
\begin{equation}
H=\frac{p_\alpha^2}{2\mu_\alpha}+\frac{q_\alpha^2}{2M_\alpha}+\sum_{i=1}^3v_i,
\label{3hamilt}
\end{equation}
with $\alpha\equiv 1,2,3$. Here $v_1$ is the interaction between particles 2 and 3, $p_1$ is the relative momentum of particles 2 and 3 and $q_1$ the momentum of particle 1 with respect to the center of mass of particles 2 and 3. We are using the odd-man-out notation where the index indicates the particle that is not being considered. The reduced masses are given by $\mu_1\equiv\frac{m_2m_3}{m_2+m_3}$ and $M_1\equiv\frac{m_1(m_2+m_3)}{m_1+m_2+m_3}$. 

Eq. (\ref{3hamilt}) allows us to define now a two-body Hamiltonian, $h$, a two-body resolvent, $g$, and a two-body $T$ operator, $t$, associated to the two-body channels, as
\begin{eqnarray}
&&h_\alpha=\frac{p_\alpha^2}{2\mu_\alpha}+\frac{q_\alpha^2}{2M_\alpha}+v_\alpha,\\
&&g_\alpha(z)\equiv(z-h_\alpha)^{-1},\\
&&t_\alpha(z)=v_\alpha+v_\alpha g_\alpha(z)v_\alpha.
\end{eqnarray}
The full hamiltonian can be written as
\begin{equation}
H=h_\alpha+\bar{v}_\alpha\;\;\;{\rm with}\;\;\;\bar{v}_\alpha=\sum_{i=1}^3v_i-v_\alpha.
\end{equation}

However, the existence of multiple channels brings another problem. Consider the full Green function ($G(z)\equiv (z-H)^{-1}$) written as the following identity
\begin{eqnarray}
\nonumber
&&g_\alpha^{-1}(z)-G^{-1}(z)=z-h_\alpha-z+H\\
&&G(z)=g_\alpha(z)+g_\alpha(z)\bar{v}_\alpha G(z)=g_\alpha(z)+ G(z)\bar{v}_\alpha g_\alpha(z),
\end{eqnarray}
where the second line is obtained multiplying conveniently the first line by $G(z)$ and $g_\alpha(z)$. Then, the Lippmann-Schwinger equation for scattering can be obtained replacing the above equation into Eq. (\ref{hom}):
\begin{equation}
|\Psi_\alpha^{(\pm)}\rangle=\lim_{\epsilon\to0^+}\pm i\epsilon G(E\pm i\epsilon)|\Phi_\alpha\rangle,
\end{equation}
in such way that 
\begin{equation}
|\Psi_\alpha^{(\pm)}\rangle=\lim_{\epsilon\to0^+}\pm i\epsilon\left[g_\alpha(E\pm i\epsilon)+
g_\alpha(E\pm i\epsilon)\bar{v}_\alpha G(E\pm i\epsilon)\right]|\Phi_\alpha\rangle,
\label{phia}
\end{equation}
as $\lim_{\epsilon\to0^+}\pm i\epsilon\,g_\alpha(E\pm i\epsilon)|\Phi_\alpha\rangle=|\Phi_\alpha\rangle$, we have
\begin{equation}
|\Psi_\alpha^{(\pm)}\rangle=|\Phi_\alpha\rangle+g_\alpha(E\pm i\epsilon)\bar{v}_\alpha |\Psi_\alpha\rangle^{(\pm)}.
\end{equation}
The above scattering equation, where may exists more than one channel, is not uniquely determined as the homogeneous part presents a different solution of the trivial one as follows. Consider another state $|\Phi_\beta\rangle$ which is not an eigenvalue of $h_\alpha$. If we replace $\Phi_\alpha$ by $\Phi_\beta$ in Eq. (\ref{phia}), we have in this case that $\lim_{\epsilon\to0^+}\pm i\epsilon g_\alpha(E\pm i\epsilon)|\Phi_\beta\rangle=0$ and consequently:
\begin{equation}
|\Psi_\beta^{(\pm)}\rangle=g_\alpha(E)\bar{v}_\alpha |\Psi_\beta\rangle^{(\pm)},
\end{equation}
then the homogeneous part has non-trivial solutions (any multiple of this equation is also a solution). As a conclusion, this part inserted in the inhomogeneous equation returns infinite solutions. 

The problem of non-uniqueness of Lippmann-Schwinger equations when considering three or more particles was solved in 1960 by L. D. Faddeev \cite{faddeev}. The idea of Faddeev is the following (we are not intending to go deep in the mathematical aspects). Let us start separating the potential in three terms, $v_k$, representing the pairwise interactions between particles $ij$ ($i,j,k=1,2,3$) and defining the operators $T_i$ as:
\begin{equation} 
T_i\equiv v_i+v_iG_0T,
\label{split}
\end{equation}
in such a way that the full transition operator is given by $T_1+T_2+T_3$ ($V=v_1+v_2+v_3$). We can write this equation in a matrix form as
\begin{equation}
\left( \begin{array} [h] {l}
T_1 \\ T_2 \\ T_3 \end{array} \right)=
\left( \begin{array} [h] {l}
v_1 \\ v_2 \\ v_3 \end{array} \right)+
\left( \begin{array}{*{3}{c@{\;}}}
v_1 & v_1 & v_1 \\
v_2 & v_2 & v_2 \\
v_3 & v_3 & v_3 \end{array}\right)
G_0
\left( \begin{array} [h] {l}
T_1 \\ T_2 \\ T_3 \end{array} \right).
\label{matrixT}
\end{equation}
Eq. (\ref{matrixT}) still presents a non-compact kernel ($\not\in \mathcal{L}^2$), which cannot be solved by any iteration. You can iterate as many times as you want that you will always have terms where one particle do not interact with the others resulting in the appearance of delta functions (see \cite{gloeckle} for a discussion about disconnected diagrams). However, a simple manipulation of Eq. (\ref{split}) seems to deal positively with the singularity of the kernel. Separating the terms of the full $T$ on the right hand side of Eq. (\ref{split}), we can write it as:
\begin{eqnarray}
&&(1-v_1G_0)T_1=v_1+v_1G_0(T_2+T_3)\\
&&T_1=t_1+t_1G_0(T_2+T_3), 
\label{split1}
\end{eqnarray}
where $t_1={(1-v_1G_0)}^{-1}v_1$ is a two-body operator. Clearly, the same procedure can be made for the other components. Thus, Eq. (\ref{matrixT}) can now be written as  
\begin{eqnarray}
\left( \begin{array} [h] {l}
T_1 \\ T_2 \\ T_3 \end{array} \right)=
\left( \begin{array} [h] {l}
t_1 \\ t_2 \\ t_3 \end{array} \right)+
\left( \begin{array}{*{3}{c@{\;}}}
0 & t_1 & t_1 \\
t_2 & 0 & t_2 \\
t_3 & t_3 & 0 \end{array} \right)G_0
\left( \begin{array} [h] {l}
T_1 \\ T_2 \\ T_3
\end{array} \right).
\label{faddeeveq}
\end{eqnarray}

The set of equations, written in a matrix form in (\ref{faddeeveq}), is known as Faddeev equations for the three-body $T$-operator. Note that the kernels of these equations still present delta functions
\begin{equation}
\langle\vec{p}_i,\vec{q}_i|t_i(z)|\vec{p}_i\,^\prime,\vec{q}_i\,^\prime\rangle=\delta(\vec{q}_i-\vec{q}_i\,^\prime)
\langle\vec{p}_i|t_i(z-q_i^2/2M_i)|\vec{p}_i\,^\prime\rangle,
\end{equation}
and, consequently, the kernel is not square integrable. But, differently from Eq. (\ref{matrixT}), it is possible to disappear with the delta functions after a first iteration as follows:
\begin{eqnarray}
&&\left( \begin{array} [h] {l}
T_1 \\ T_2 \\ T_3 \end{array} \right)=
\left( \begin{array} [h] {l}
t_1 \\ t_2 \\ t_3 \end{array} \right)+
\left( \begin{array}{*{3}{c@{\;}}}
0 & t_1 & t_1 \\
t_2 & 0 & t_2 \\
t_3 & t_3 & 0 \end{array} \right)G_0
\left\{\left( \begin{array} [h] {l}
t_1 \\ t_2 \\ t_3 \end{array} \right)+
\left( \begin{array}{*{3}{c@{\;}}}
0 & t_1 & t_1 \\
t_2 & 0 & t_2 \\
t_3 & t_3 & 0 \end{array} \right)
\left( \begin{array} [h] {l}
T_1 \\ T_2 \\ T_3
\end{array} \right)\right\} \\ \nonumber
&&=\left( \begin{array} [h] {l}
t_1 \\ t_2 \\ t_3 \end{array} \right)+
\left( \begin{array} [h] {l}
t_1G_0(t_2+t_3) \\ t_2G_0(t_1+t_3) \\ t_3G_0(t_1+t_2) \end{array} \right) \\
&&+
\left( \begin{array}{*{3}{c@{\;}}}
t_1G_0(t_2+t_3) & t_1G_0t_3 & t_1G_0t_2 \\
t_2G_0t_3 & t_2G_0(t_1+t_3) & t_2G_0t_1 \\
t_3G_0t_2 & t_3G_0t_1 & t_3G_0(t_1+t_2) \end{array} \right)G_0
\left( \begin{array} [h] {l}
T_1 \\ T_2 \\ T_3
\end{array} \right).
\label{faddeeviter}
\end{eqnarray}
Note that now all the terms in the kernel are of this form $t_iG_0t_j$ with $i\neq j$ in such a way the three particles are always linked together and the delta function is no longer present (all diagrams are now connected \cite{gloeckle}). As a consequence, the kernel belongs to the Hilbert-Schmidt class and the Fredholm method can be applied and the Faddeev equation has a unique solution.

\section{Bound and scattering states}

In this section we start to apply the scattering theory showed ``en passant'' in the last section. As anticipated in the title of these lectures, everything here will be calculated in momentum space. We also have to call the attention that we will focus only in universal situations, i.e., the range of the potential is much smaller than the typical sizes of the system in a manner that the observables do not depend on the form of the short-range potential.

\subsection{Two-body $T$-operator for a separable potential}

A potential is called separable if 
\begin{equation}
\langle\vec{p}|V|\vec{p}\,^\prime\rangle=\lambda g(\vec{p})g^\star(\vec{p}\,^\prime).
\end{equation}
We can write generically a rank-1 separable potential in the form of a projection operator as
\begin{equation}
V\equiv\lambda|\chi\rangle\langle\chi|,
\label{seppot}
\end{equation}
where $\lambda$ is the strength of the potential and $\langle\vec{p}\,|\chi\rangle=g(p)$ and $\langle\chi|\vec{p}\,\rangle=g^*(\vec{p})$ are the form factors.

Let us start replacing the separable potential, written in the form of Eq. (\ref{seppot}), in the $T$-operator equation

\begin{eqnarray}
&&t(E)=V+VG_0(E)t(E), \label{eqsub21} \\ \nonumber \\ \label{eqsub21a}
&&t(E)=\lambda|\chi\rangle\langle\chi|+\lambda|\chi\rangle\langle\chi|G_0(E)t(E),
\end{eqnarray}
multiplying by $\langle\chi|G_0(E)$ from the left side and isolating $\langle\chi|G_0(E)t(E)$ we have:
\begin{eqnarray}
\nonumber
&&\langle\chi|G_0(E)t(E)=\lambda\langle\chi|G_0(E)|\chi\rangle\langle\chi|+
\lambda\langle\chi|G_0(E)|\chi\rangle\langle\chi|G_0(E)t(E), \\ \nonumber \\ \nonumber
&&(1-\lambda\langle\chi|G_0(E)|\chi\rangle)\langle\chi|G_0(E)t(E)=
\lambda\langle\chi|G_0(E)|\chi
\rangle\langle\chi|, \\ \nonumber \\
&&\langle\chi|G_0(E)t(E)=\frac{\lambda\langle\chi|G_0(E)|\chi\rangle\langle\chi|}
{1-\lambda\langle\chi|G_0(E)|\chi\rangle}.
\label{eqsub22}
\end{eqnarray}
Replacing Eq. (\ref{eqsub22}) in (\ref{eqsub21}) we get the two-body $T$-operator:
\begin{eqnarray}
\nonumber
&&t(E)=\lambda|\chi\rangle\langle\chi|+\frac{\lambda^2|\chi\rangle\langle
\chi|G_0(E)|\chi
\rangle\langle\chi|}{1-\lambda\langle\chi|G_0(E)|\chi\rangle}, \\ \nonumber \\ 
&&t(E)=\lambda|\chi\rangle\left( 1+\frac{\lambda\langle\chi|G_0(E)|\chi\rangle}
{1-\lambda\langle\chi|G_0(E)|\chi\rangle} \right) \langle\chi|.
\end{eqnarray}
Then, we can write it as:
\begin{equation}
t(E)=|\chi\rangle\tau(E)\langle\chi|,
\label{eqsub23}
\end{equation}
Note here that a separable potential results in a separable $T$-operator. The function $\tau$ is given by:
\begin{equation}
\tau(E)=\frac{1}{\lambda^{-1}-\langle\chi|G_0(E)|\chi\rangle}.
\label{eqsub24}
\end{equation}

Writing explicitly the matrix element of Eq. (\ref{eqsub23}), we have that:

\begin{equation}
\tau(E)={\left( \lambda^{-1}-\int d^3p\frac{|g(\vec{p})|^2}
{E-\frac{p^2}{2M}+i\epsilon}\right)}^{-1},
\label{eqsub25}
\end{equation}

\noindent
where $M$ is the reduced mass of the two-body system.

Since now,  we just assumed that the two-body potential is separable. We didn't give any other information about it. However, from this point we will restrict it a little more. We will focus here only on the universal characteristics of the system. The universal characteristics appear in systems that present an important property called {\it universality}. In these systems the calculated observables do not depend on the details of the short-range potential. This peculiar situation occurs when the size of the system, represented by a typical scale (the two-body scattering length, for example), is much larger than the range of the potential. This situation can be achieved by construction using a Dirac-delta potential. As this potential has a range equal to zero, any size of the system is infinitely larger than the range of the potential, then any quantity calculated is universal by definition. For a zero-range potential $V(\vec{r})=\lambda\delta(\vec{r})$ the form factor $g(\vec{p})=\langle\vec{p}\,|\chi\rangle=1$.

\noindent\rule[0.5ex]{\linewidth}{1pt}
{\bf Exercise 5:} Starting from the definition of a local potential, 
\begin{equation}
\nonumber
\langle\vec{r}\,^\prime|V|\vec{r}\rangle=\delta(\vec{r}\,^\prime-\vec{r})V(\vec{r}),
\end{equation}
show that the form factor for a Dirac-delta potential is equal to 1. Show also that this potential is separable.

\noindent\rule[0.5ex]{\linewidth}{1pt}

The consequence of replacing $g(\vec{p})=1$ is that the integral in
\begin{equation}
\tau^{-1}(E)=\lambda^{-1}-\int d^3p\frac{1}{E-\frac{p^2}{2M}+i\epsilon}
\label{tauzerorange}
\end{equation}
diverges for large momenta. In order to transform Eq. (\ref{tauzerorange}) into a finite value we need that $\lambda^{-1}$ should also be infinite. Calculating $\tau^{-1}(-|E_2|)$ at the two-body binding energy, $E_2$, and remembering that a bound state is a pole in the $T$-operator we have that:
\begin{eqnarray}
&&\tau^{-1}(-|E_2|)=0=\lambda^{-1}-\int d^3p\frac{1}{-|E_2|-\frac{p^2}{2M}}\\
&&\lambda^{-1}=\int d^3p\frac{1}{-|E_2|-\frac{p^2}{2M}},
\label{renorm}
\end{eqnarray}
then we can replace $\lambda^{-1}$ in Eq. (\ref{tauzerorange}) (note that $\lambda^{-1}$ also diverges) obtaining:
\begin{eqnarray}
\tau^{-1}(E)=-\int d^3p\left(\frac{1}{|E_2|+\frac{p^2}{2M}}+\frac{1}{E-\frac{p^2}{2M}+i\epsilon}\right).
\label{taurenorm}
\end{eqnarray}
Now the integral is finite and can be calculated using the residue theorem giving:
\begin{equation}
\tau^{-1}(E)=2\pi^2(2M)^{3/2}\left(\sqrt{|E_2|}-\sqrt{|E|}\right).
\label{taufinal}
\end{equation}
Note that the introduction of the physical scale $E_2$ not only regularizes the integral but also renormalizes it.

\noindent\rule[0.5ex]{\linewidth}{1pt}
{\bf Exercise 6:} Calculate explicitly Eq. (\ref{taufinal}). Remember that the energy can be written as $E=k^2/(2M)$ and use the residue theorem. $k=\pm i\sqrt{2M|E|}$ where the plus sign is related to a two-body bound state and the minus to a virtual. The states of negative energy are located at the first (physical) and second (unphysical) sheets. 

\noindent\rule[0.5ex]{\linewidth}{1pt}

\noindent\rule[0.5ex]{\linewidth}{1pt}
{\bf Exercise 7:} Show that the Lippmann-Schwinger equation given by Eq. (\ref{LS}) can be written as (we dropped out the sub and superscripts)
\begin{equation}
\nonumber
|\Psi\rangle=|\Phi\rangle+G_0T|\Phi\rangle
\end{equation}

\noindent\rule[0.5ex]{\linewidth}{1pt}

\noindent\rule[0.5ex]{\linewidth}{1pt}
{\bf Exercise 8:} Starting from the Lippmann-Schwinger equation written in the form given in the last exercise, show that the scattering amplitude $f(\theta,\phi)$ is related to the $T$-operator as 
\begin{equation}
\nonumber
f(\theta,\phi)=-\frac{1}{4\pi}\langle\vec{k}\,^\prime|T|\vec{k}\,^\prime\rangle
\end{equation}

\noindent\rule[0.5ex]{\linewidth}{1pt}

\subsection{Three-body $T$-operator for a separable potential}

We saw that the regularization (and renormalization) of Eq. (\ref{tauzerorange}) appears naturally after calculating $\tau$ at a bound state energy and noting that such energy is a pole in the $T$-operator. However, the choice of the two-body energy $|E_2|$ as a scale is arbitrary - it could be another observable or we could just put a cuttoff for high momenta and further relate it with some observable. 

The three-body $T$-operator may also be regularized in a similar way. Let us start defining the three-body $T$-operator at a given energy $E=-\mu^2$
\begin{eqnarray}
\nonumber
&&T(-\mu^2)=\left[1+T(-\mu^2)G_0(-\mu^2)\right]V, \\ 
&&V=\left[(1+T(-\mu^2)G_0(-\mu^2))\right]^{-1}T(-\mu^2),
\label{eqsub31}
\end{eqnarray}
replacing the above $V$ into $T(E)=V+VG_0^{(+)}(E)T(E)$ we get:

\begin{equation}
T_R(E,\mu^{2})=T_R(-\mu^{2})+T_R(-\mu^{2})\left(G_0^{(+)}(E)-G_0(-\mu^2)\right)T_R(E),
\label{eqsub32}
\end{equation}
where we inserted the subscript $R$ to indicate a regularized (renormalized) $T$-operator. 

\noindent\rule[0.5ex]{\linewidth}{1pt}
{\bf Exercise 9:} Iterate Eq. (\ref{eqsub32}) and show that it can be written as 
\begin{equation}
T_R^{(m)}(E,\mu^{2})=T_R(-\mu^2)\sum_{n=0}^{m-1}\left[\left(G_0^{(+)}(E)-
G_0(-\mu^2)\right)T_R(-\mu^2)\right]^n.
\label{eqsub33}
\end{equation}
Note that truncating the series at the $m+1$ element (the term that contains $T(E)$) the kernel subtraction appears in all intermediate terms until $m-1$.

\noindent\rule[0.5ex]{\linewidth}{1pt}

\noindent\rule[0.5ex]{\linewidth}{1pt}
{\bf Exercise 10:} Take the derivative of Eq. (\ref{eqsub33}) with respect to $\mu^2$ and show that the results obtained from the subtracted equation do not depend on the choice of the subtraction point, $\mu^2$.
\begin{equation}
\frac{d}{d\mu^2}T_R(E)=0
\label{eqsub36}
\end{equation}

\noindent\rule[0.5ex]{\linewidth}{1pt}

As showed in the last exercise $T(E,\mu^2)$ does not depends on the subtraction point $\mu^2$ in such a way we will write the $T$-operator simply as
\begin{equation}
T_R(E)=T_R(-\mu^{2})+T_R(-\mu^{2})\left(G_0^{(+)}(E)-G_0(-\mu^2)\right)T_R(E).
\label{eqsub32a}
\end{equation}

Note that Eq. (\ref{eqsub32a}) has the same operatorial form as the original equation for the $T$-operator (this can be seen making the following replacements $T_R(-\mu^{2})\equiv V(-\mu^2)$ and $\left(G_0^{(+)}(E)-G_0(-\mu^2)\right)\equiv G_0(E;-\mu^2)$). Let us assume an ansatz and define the three-body $T$-operator in the subtraction point as the sum of all pairs of the renormalized two-body $T$-operators ($t_{R_\alpha}$)

\begin{equation}
T_{R}(-\mu_{(3)}^2)=\sum_{\alpha,\beta,\gamma}t_{R_\alpha}\left(-\mu_{(3)}^{2}-
\frac{q_{\alpha}^2}{2m_{\beta\gamma,\alpha}}\right),
\label{eqsub38}
\end{equation}
where $\alpha,\beta,\gamma=1,2,3$ ($\alpha\neq\beta\neq\gamma$) and $m_{\beta\gamma,\alpha}$ is the reduced mass of the pair $\beta\gamma$ and $\alpha$. Note that the argument of $t_{R}$ is the energy of the center of mass of the pair. From this point the two and three-body $T$-operators will be represented, respectively, by $t$ and $T$. The counterpart of regularized/renormalized two and three-body $T$-operators is the addition of two physical scales which will be generically represented by $\mu_{(i)}^{2}$, where the subscript $i=2,3$ distinguishes between the two and three-body scales. The original $T$-operator is recovered in the limit $\mu\rightarrow\infty$.

Replacing Eq. (\ref{eqsub38}) in (\ref{eqsub32a}) we have:

\begin{equation}
T_R(E)=\sum_{\alpha,\beta,\gamma}t_{R_\alpha}\left(-\mu_{(3)}^{2}-
\frac{q_{\alpha}^2}{2m_{\beta\gamma,\alpha}}\right)
\left[1+\left(G_0^{(+)}(E)-G_0(-\mu_{(3)}^2)\right)T_R(E)\right].
\label{eqsub39}
\end{equation}

Defining the component 1 of the three-body $T$-operator as
\begin{equation}
T_{R_1}(E)=t_{R_1}\left(-\mu_{(3)}^{2}-
\frac{q_{1}^2}{2m_{23,1}}\right)\left[1+\left(G_0^{(+)}(E)-
G_0(-\mu_{(3)}^2)\right)T_R(E)\right],
\label{eqsub40}
\end{equation}
and writing $T_R(E)$ as the sum of the components corresponding to an interacting pair (we are using the odd-man-out notation where the subscript 1 says that the pair $(23)$ interacts, 2 $(13)$ and 3$(12)$) as
\begin{equation}
T_{R}(E)=\sum_{\alpha=1,2,3}T_{R_{\alpha}}(E),
\label{eqsub40a}
\end{equation}
we may finally write the Faddeev component 1 as a function of the 
other two:

\begin{equation}
T_{R_1}(E)=t_{R_1}\left(E-
\frac{q_{1}^2}{2m_{23,1}}\right)\left[1+\left(G_0^{(+)}(E)-
G_0(-\mu_{(3)}^2)\right)\left(T_{R_2}(E)+T_{R_3}(E)\right)\right],
\label{eqsub41}
\end{equation}

\noindent\rule[0.5ex]{\linewidth}{1pt}
{\bf Exercise 11:} Verify Eq. (\ref{eqsub41}). 

\noindent\rule[0.5ex]{\linewidth}{1pt}

\subsection{Bound state equations}

Consider the following completeness relation:
\begin{equation}
\mathbf1=\sum_B|\Phi_B\rangle\langle\Phi_B|+\int d^3k|\Psi_{k}^{(+)}
\rangle\langle\Psi_{k}^{(+)}|,
\label{ev1}
\end{equation} 
\noindent
where the first and second terms containing $|\Phi_B\rangle$ and $|\Psi_k^{(+)}\rangle$, decomposes, respectively, a given state into bound and scattering states of initial momentum $\vec{k}$. Inserting Eq. (\ref{ev1}) in the $T$-operator equation we have:

\begin{eqnarray}
\nonumber
&&T(E)=V+\sum_B VG^{(+)}(E)|\Phi_B\rangle\langle\Phi_B|V+
\int d^3kVG^{(+)}(E)|\Psi_{k}^{(+)}\rangle\langle\Psi_{k}^{(+)}|V; 
\\ \nonumber \\
&&T(E)=V+\sum_B\frac{V|\Phi_B\rangle\langle\Phi_B|V}{E-E_B+i\epsilon}
+\int d^3k\frac{V|\Psi_{k}^{(+)}\rangle\langle\Psi_{k}^{(+)}|V}
{E-{\bar E}_k+i\epsilon},
\label{ev2}
\end{eqnarray}
where the complete propagator $G^{(+)}(E)$ was explicitly written in terms of the bound ($E_B<0$) and continuous (${\bar E}_k>0$) state eigenvalues. The $T$-operator written in the form of Eq. (\ref{ev2}) is called Low equation \cite{LoPR55}. For a given bound states $(E\approx E_B)$ we have that the second term is dominant due to the presence of a pole. Then, we can write:
\begin{equation}
T(E)\approx\frac{V|\Phi_B\rangle\langle\Phi_B|V}{E+|E_B|}
=\frac{|\Gamma_B\rangle\langle\Gamma_B|}{E+|E_B|},
\label{ev3}
\end{equation}
where we defined $|\Gamma_B\rangle=V|\Phi_B\rangle$. The same applies to a Faddeev component of the $T$-operator:

\begin{eqnarray}
&&T_\alpha=v_\alpha + v_\alpha GV\\
&&T_\alpha(E)\approx\frac{v_\alpha|\Phi_\alpha\rangle\langle\Phi_\alpha|V}{E+|E_\alpha|}
=\frac{|\Gamma_\alpha\rangle\langle\Gamma|}{E+|E_\alpha|},
\label{ev4}
\end{eqnarray}
with $\alpha=1,2,3$, $|\Gamma_\alpha\rangle=v_\alpha|\Phi_\alpha
\rangle$ and $\langle\Gamma|=\langle\Phi_\alpha|V$. Replacing Eq. 
(\ref{ev4}) in (\ref{eqsub41}), we have:

\begin{eqnarray}
\nonumber
&&\frac{|\Gamma_1\rangle\langle\Gamma|}{E+|E_1|}\approx
t_{R_1}\left(E-\frac{q_{1}^2}{2m_{23,1}}\right)
\left\{1+\left[G_{0}^{(+)}(E)-G_0(-\mu_{(3)}^2)\right]\right. 
\\ \nonumber \\ 
&&\times\left.\left(\frac{|\Gamma_2
\rangle\langle\Gamma|}{E+|E_1|}+\frac{|\Gamma_3
\rangle\langle\Gamma|}{E+|E_1|}\right)\right\},
\label{ev5}
\end{eqnarray}
where $E_1$ is the energy of the bound pair $23$. Cancelling the common terms on both sides we finally have the homogeneous equation

\begin{equation}
|\Gamma_1\rangle=t_{R_1}\left(E-\frac{q_{1}^2}{2m_{23,1}}\right)
\left(G_{0}^{(+)}(E)-G_0(-\mu_{(3)}^2)\right)\left(|\Gamma_2\rangle+|\Gamma_3\rangle\right).
\label{ev6}
\end{equation}
Remember that this approximation is as good as closer to the limit $E\rightarrow -|E_1|$.

Writing explicitly the two-body $T$-operator in the operatorial form given by Eq. (\ref{eqsub23}):
\begin{equation}
|\Gamma_1\rangle=|\chi\rangle\tau\left(E-\frac{q_{1}^{2}}{2m_{23,1}}\right)
\langle\chi|\left(G_{0}^{(+)}(E)-G_0(-\mu_{(3)}^{2})\right)
\left(|\Gamma_2\rangle+|\Gamma_3\rangle\right),
\label{ev7}
\end{equation}
\noindent
where $\tau(E)$ is the function given by Eq. (\ref{taufinal}). Multiplying Eq. (\ref{ev7}) by $\langle\vec{p}_1,\vec{q}_1|$ 
from the left, we get:

\begin{equation}
\langle\vec{p}_1,\vec{q}_1|\Gamma_1\rangle=\langle\vec{p}_1|\chi\rangle
\tau\left(E-\frac{q_{1}^{2}}{2m_{23,1}}\right)\langle\chi|\langle\vec{q_1}|
\left(G_{0}^{(+)}(E)-G_0(-\mu_{(3)}^{2})\right)
\left(|\Gamma_2\rangle+|\Gamma_3\rangle\right),
\label{ev8}
\end{equation}
using that 
\begin{equation}
\langle\vec{p}_1,\vec{q}_1|\Gamma_1\rangle=\langle\vec{p}_1,\vec{q}_1|V|\Phi_1\rangle=
\int d\vec{p}\,^\prime\langle\vec{p}_1|\chi\rangle\langle\chi|\vec{p}\,^\prime\rangle\langle\vec{q}_1,\vec{p}\,^\prime|\Phi_1\rangle=
f_1(\vec{q_1})
\end{equation}
we have the homogeneous equation for a three-body bound state for the Faddeev component 1. 

\begin{equation}
f_1(\vec{q_1})=\langle\vec{p}_1|\chi\rangle\tau\left(E-\frac{q_{1}^{2}}{2m_{23,1}}\right)\langle\chi|\langle
\vec{q_1}|\left(G_{0}^{(+)}(E)-G_0(-\mu_{(3)}^{2})\right)
\left(|\Gamma_2\rangle+|\Gamma_3\rangle\right).
\label{ev9}
\end{equation}
The function $f$ is called {\it spectator function} \cite{mitra}. Note that for the specific case of three identical bosons, the three spectator functions are exactly the same, such that:
\begin{equation}
\langle\vec{q}_1|f_1\rangle=
\langle\vec{q}_2|f_2\rangle=
\langle\vec{q}_3|f_3\rangle.
\label{ev10}
\end{equation}
Then, we have now to calculate the matrix elements on the right side of Eq. (\ref{ev9}). 

\noindent\rule[0.5ex]{\linewidth}{1pt}
{\bf Exercise 12:} Show that the matrix element $\langle\chi|\langle
\vec{q_1}|G_{0}^{(+)}(E)|\Gamma_2\rangle$ is given by

\begin{equation}
\int d^3q_2^\prime\frac{1}{E-q_1^2-{q_2}^2-\vec{q}_1\cdot\vec{q}_2}f(\vec{q}_2),
\label{melement}
\end{equation}
where we already replaced the form factors for the Dirac-delta potential 
$\langle\vec{p}|\chi\rangle=g(\vec{p})=1$ and the masses $m_1=m_2=m_3=1$.

\begin{figure}[htb!]
\centering
\includegraphics[width=10cm]{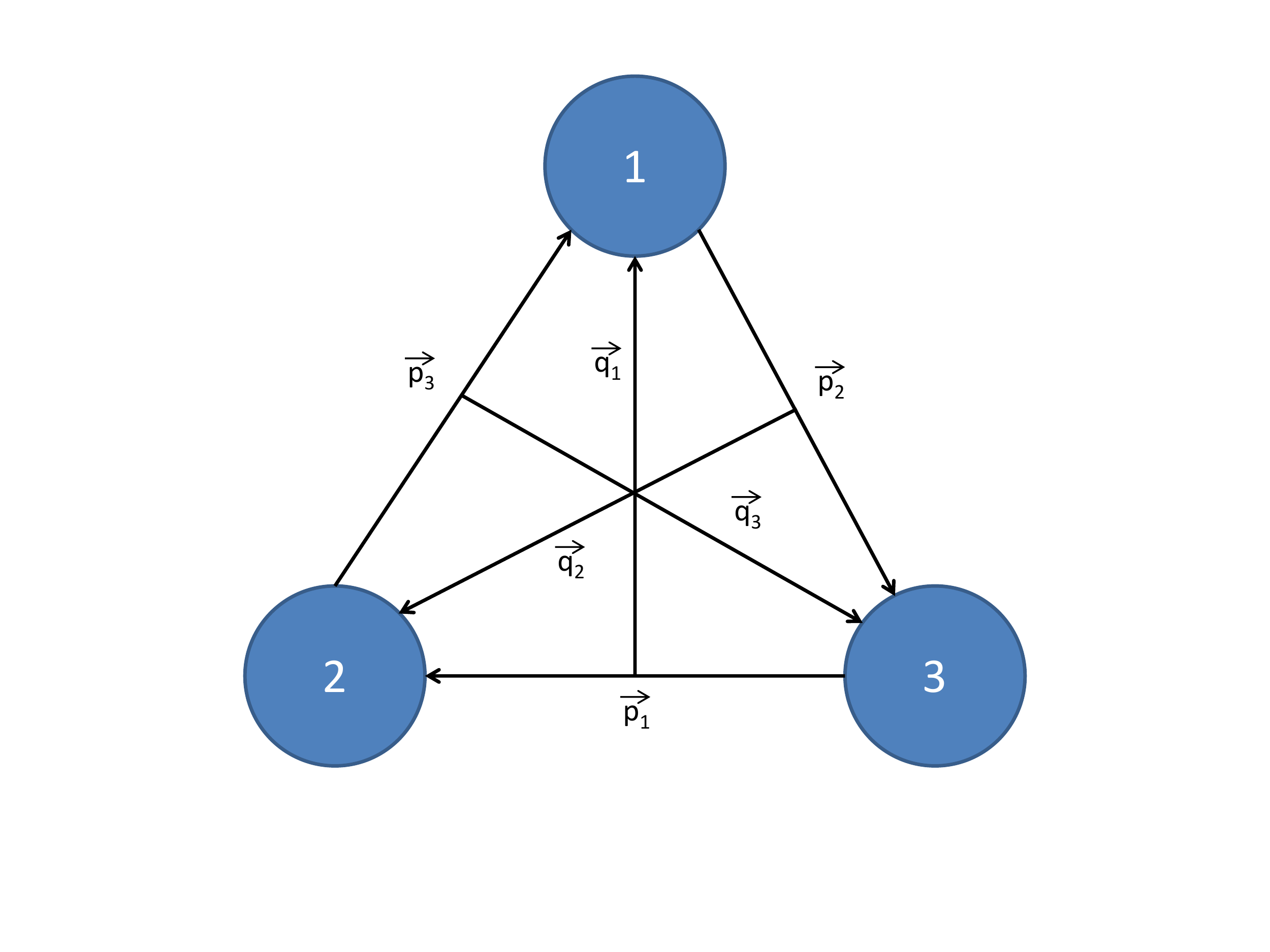}
\caption{Jacobi coordinates.}
\label{figjacobi}
\end{figure}

To calculate Eq. (\ref{melement}) you will need to use the relations between 
Jacobi coordinates (the relative coordinates between the constituents of the 
system) showed in Fig. \ref{figjacobi} for a generic three-body system.

The momenta of the particles with respect to the center of mass satisfy the relation $\vec{k}_i+\vec{k}_j+\vec{k}_k=0$. Denoting by $m_\alpha$ and $v_\alpha$ ($\alpha=1,2,3$) their masses and velocities, we may deduce that $\vec{p}_1$ e $\vec{q}_1$ are given by:

\begin{eqnarray}
&&\vec{p}_1=\frac{m_2m_3}{m_2+m_3}(\vec{v}_2-\vec{v}_3)=
\frac{m_3\vec{k}_2-m_2\vec{k}_3}{m_2+m_3}\\
&&\vec{q}_1=\frac{m_1(m_2+m_3)}{m_1+m_2+m_3}\left[\vec{v}_1-
\frac{m_2\vec{v}_2+m_3\vec{v}_3}{m_2+m_3}\right]=\vec{k}_1.
\end{eqnarray}
$\vec{q}_j=\vec{k}_j$ and $\vec{q}_k=\vec{k}_k$ are defined analogously and may be found by cyclic permutation of the indexes. Then, combining the coordinates we have that:
\begin{eqnarray}
&&\vec{p}_1=\frac{m_3\vec{q}_2-m_2\vec{q}_3}{m_2+m_3}\\
&&\vec{p}_2=\frac{m_1\vec{q}_3-m_3\vec{q}_1}{m_1+m_3}\\
&&\vec{p}_3=\frac{m_2\vec{q}_1-m_1\vec{q}_2}{m_1+m_2}
\end{eqnarray}

\noindent\rule[0.5ex]{\linewidth}{1pt}

Using the result from Eq. (\ref{melement}) we may write the homogeneous equation for three identical bosons:

\begin{eqnarray}
\nonumber
&&f(\vec{q}\,)=2\tau\left(-|E_3|-\frac{3}{4}q^2\right) \\
&&\times\int d^3q^\prime \left(\frac{1}{-|E_3|-q^2-{q^\prime}^2-\vec{q}\,^\prime\cdot\vec{q}}
-\frac{1}{-\mu_{(3)}^2-q^2-{q^\prime}^2
-\vec{q}\,^\prime\cdot\vec{q}}\right)f(\vec{q}\,^\prime),
\label{ev15}
\end{eqnarray}
where we removed all indexes. This is the Skorniakov and Ter-Martirosian (STM) equation for the bound state and zero-range potential \cite{stm}. It is worth to remind that the three-body scale $\mu_{(3)}$ is arbitrary. All momenta and $E_3$ can be rescaled with respect to $\mu_{(3)}$ in order to have dimensionless quantities as $|\epsilon_{3}|=\frac{|E_3|}{\mu_{(3)}^2}$, $y=\frac{q}{\mu_{(3)}}$ and $x=\frac{q^\prime}{\mu_{(3)}}$. Thus, the homogeneous equation (\ref{ev15}) for the three-body bound state is rewritten as
\begin{eqnarray}
\nonumber
f(y)&=&4\pi\tau\left(-|\epsilon_3|-\frac{3}{4}y^2\right)\int_0^\infty dx x^2\int_1^{-1}dz\left[
\frac{1}{|\epsilon_3|+y^2+x^2+xyz} \right. \\ \nonumber \label{ev18} \\
&&\left.-\frac{1}{1+y^2+x^2+xyz}\right]f(x),
\end{eqnarray}
where $z\equiv\cos(\vec{q}\cdot\vec{q}\,^\prime$.

Here, we might be tempted to use the solution coming from the Fredholm theory, however, we have here two problems: we don't know neither $f$ nor $\epsilon_3$. Among several numerical methods that we can use to solve this problem, we will focus in only one. Generically, the structure of this integral equation reads (after integrating out the angular part) 
\begin{equation}
f(y)=\int dx K(y,x;E)f(x).
\label{homgen}
\end{equation}
In order to calculate this equation numerically we should discretize it. Let us call by $f_i\equiv f(y_i)$ the value of $f$ calculated in a given mesh point $y_i\,\,\,(i=1,\dots,N)$. We then have
\begin{eqnarray}
&&f(y_i)=\sum_{j=1}^N wx_j K(y_i,x_j;E)f(x_j)\\
&&f_i=\sum_{j=1}^N w_j K_{ij}(E)f_j\\
&&\sum_{j=1}^N\left(\delta_{ij}-w_jK_{ij}(E)\right)f_j=0\rightarrow MF=0,
\end{eqnarray}
where $K(y_i,x_j;E)\equiv K_{ij}(E)$, $wx_i\equiv w_i$ is a given weight associated to the mesh point $x_i$ (if you are not familiar with these terms, search for Gauss-Legendre quadrature), and $\delta_{ij}$ is a Kronecker delta. We then have a homogeneous equation with matrices $M$ and $F$ given by:
\begin{equation}
M=
\left( \begin{array} {*{4}{c@{\;\;}}}
1-w_1K_{11}(E) & -w_2K_{12}(E) & \cdots & -w_NK_{1N}(E) \\
-w_1K_{21}(E) & 1-w_2K_{22}(E) & \cdots & -w_NK_{2N}(E) \\
\vdots & \vdots & \ddots & \vdots \\
-w_1K_{N1}(E) & -w_2K_{N2}(E) & \cdots & -w_NK_{NN}(E) \\
\end{array} \right)\,\,{\rm and}\,\,
F=
\left( \begin{array}[h]{l}
f_1 \\ 
f_2 \\ 
\vdots\\
f_N
\end{array} \right).
\label{eqsub12}
\end{equation}
Now, if we want a different solution from the trivial one we have to find an $E=E_3$ that gives a determinant $\det(M(E_3))=0$. Once a bound state energy is determined, we can now discuss how to determine the three-body wave function. 

The three-body wave function may be written as a function of the spectator function. So, let us first consider how to determine $f$. As the determinant is equal to zero, one of the equations of our homogeneous system is redundant and it can be eliminated. Then, $f_1$, for example (it could be another position than 1), can be set arbitrarily as $f_1\equiv1$ and the other $f$'s are calculated with respect to this $f_1$. This is not a problem at all as the wave function will be further normalized and this arbitrariness will be washed out. Thus, we have now that 

\begin{equation}
\sum_{j=1}^N M_{ij}f_j=0 \iff \sum_{j=2}^NM_{ij}f_j=-M_{i1}f_1=-M_{i1},
\end{equation}
with $i=2,\dots,N$. Representing by ${\bar M}$ the remaining matrix after eliminating the first column and line of $M$, and by $C$ the first column of $M$, $M_{i1}\,\,(i=2,\dots,N)$, without the element $M_{11}$, we have that 
\begin{equation}
\sum_{j=2}^Nf_j={\bar F}=-{\bar M}^{-1}C.
\end{equation}
Remember that we are considering only the case where the particles are identical. For two or three different particles we will also have, respectively, two or three different spectator functions. For a general three-body system with three distinct spectator functions, the three-body wave function emerges directly from the Schroedinger equation as:

\begin{eqnarray}
&&\left(H_0+\sum_{\alpha=1,2,3}\lambda_\alpha|\chi_\alpha\rangle\langle\chi_\alpha|\right)|
\Psi\rangle=E|\Psi\rangle \\
&&(E-H_0)|\Psi\rangle=\sum_{\alpha=1,2,3}\lambda_\alpha|\chi_\alpha\rangle
\langle\chi_\alpha|\Psi\rangle,
\label{raios7}
\end{eqnarray}
where the two-body separable potential was replaced by $v_\alpha=\lambda_\alpha|\chi_\alpha\rangle\langle\chi_\alpha|$ and $H_0$ is the free Hamiltonian. Multiplying Eq. (\ref{raios7}) by $\langle\vec{q}_1,\vec{p}_1|$ from the left we may write the three-body wave function in terms of the spectator functions in the coordinates $\vec{q}_1,\vec{p}_1$ as:
\begin{equation}
\langle\vec{q}_1,\vec{p}_1|\Psi\rangle=\frac{f_1(|\vec{q}_1|)+
f_2(|\vec{p}_1-\frac{\vec{q}_1}{2}|)+f_3
(|\vec{p}_1+\frac{\vec{q}_1}{2}|)}{|E_3|+H_0}.
\label{raios9}
\end{equation}

We have now the full picture to calculate the three-body binding energies and wave function.

\subsection{Scattering states equation} 

In order to write the Lippmann-Schwinger equation in momentum space, we have to insert in Eq. (\ref{ev15}) the inhomogeneous term coming from the solution of the free problem. In configuration space this term is given by a plane wave $e^{i\vec{q}\cdot\vec{r}}$ and in momentum space it can be written as $(2\pi)^{3/2}\delta(\vec{q}-\vec{k}_i)$. Here, the momentum $\vec{q}$ represents the relative momentum between the free particle and the center of mass of the bound pair. The in and outcoming momenta are given, respectively, by $k_i$ and $k_f$ are related to the total energy of a free particle and a bound pair as $E_3=-E_2+k_i^2/2m_{23,1}=-E_2+k_f^2/2m_{23,1}$. Thus, the full equation with the inhomogeneous term reads:
\begin{eqnarray}
\label{scat}
&&f(\vec{q}\,)=(2\pi)^{3/2}\delta(\vec{q}-\vec{k}_i)\\ \nonumber
&&+2\tau\left(E_3-\frac{3}{4}q^2\right)
\int d^3q^\prime \left(\frac{1}{E_3-q^2-{q^\prime}^2-\vec{q}\,^\prime\cdot\vec{q}}
-\frac{1}{-\mu_{(3)}^2-q^2-{q^\prime}^2
-\vec{q}\,^\prime\cdot\vec{q}}\right)f(\vec{q}\,^\prime),
\end{eqnarray}
with $E_3>0$. Now, we have to insert in Eq. (\ref{scat}) the boundary condition for the elastic scattering given by:
\begin{equation}
f(\vec{q}\,)\rightarrow(2\pi)^{3/2}\delta(\vec{q}-\vec{k}_i)+\frac{h(\vec{q}\,,k_i)}{E_3\pm i\epsilon-q^2},
\label{bc}
\end{equation}
where $h(\vec{q}\,,k_i)$ is the scattering amplitude. After replacing Eq. (\ref{bc}) in (\ref{scat}) we have that 
\begin{equation}
h(\vec{q}\,,k_i)={\cal V}(q,k_i;E_3)+\int d^3q^\prime \frac{{\cal V}(q,q^\prime;E_3)}{E_3\pm i\epsilon-q^{\prime2}}h(\vec{q}\,^\prime\,,k_i),
\end{equation}
where
\begin{equation}
{\cal V}(q,k_i;E_3)=2\tau(E_3-\frac34q^2)(E_3-q^2)\left(\frac{1}{E_3-q^2-k_i^2-\vec{q}\cdot\vec{k}_i}-\frac{1}{-\mu^2-q^2-k_i^2-\vec{q}\cdot\vec{k}_i}\right).
\end{equation}
Note that we are not being very precise here as there is a missing factor $(2\pi)^{3/2}$ dividing the function $h$. However, this is meaningless as we could in principle redefine another function as ${\bar h}\equiv h/(2\pi)^{3/2}$. Now, after integrating out the angular part we will arrive in a equation very similar to Eq. (\ref{homgen}), but with an inhomogeneous term 
\begin{equation}
h(x,y)=g(x,y;E)+\int dx^\prime K(x,x^\prime;E)h(x^\prime,y).
\label{inhomgen}
\end{equation}
The numerical method we used to solve this problem is very close to the one used for bound states. The difference is that now the spectrum is continuous and the energy $E$ enters as an input. Let us call by $h_{ij}=h(y_i,x_j)$, $g_{ij}=g(y_i,x_j)$ and $K_{ij}(E)=K(y_i,x_j;E)$ the values of $h$, $g$ and $K$ calculated in the mesh points $y_i,x_i\,\,\,(i=1,\dots,N)$. Then, the discretization reads:
\begin{eqnarray}
\nonumber
&&h{ij}=g_{ij}(E)+\sum_{k=1}^M w_kK_{ik}(E)h_{kj}\\
&&\sum_{k=1}^M\left(\delta_{ik}-w_kK_{ik}\right)h_{kj}=g_{ij}\rightarrow D{\cal H}=g
\end{eqnarray}
where $w_k$ is a given weight associated to the Gauss point $x_k$ and $\delta_{ik}$ is the Kronecker delta. Now, the matrix ${\cal H}$ returns the function $h$ which is directly associated with the differential cross section $\frac{d\sigma}{d\Omega}=|h(\vec q,k_i)|^2$.

\section{Applications}

The last part of the lectures notes will be reserved to some practical examples. 

\subsection{Efimov states}

In 1970, Vitaly Efimov published a paper \cite{efimov} where he studied the energy spectrum of a system formed by bound states of three-identical bosons, interacting by a two-body short-range potential. He observed a very curious behaviour: the number of three-body bound states increased to infinity if the two-body binding energy tends to zero. It took more than 30 years to have an experimental evidence of this counter-intuitive phenomenon in the context of ultracold atoms \cite{grimm}. These bound states still present a very interesting scaling: the ratio between two consecutive bound states is given by $E_3^{(N)}/E_3^{(N+1)}=e^{2\pi/s}\,\,\,(N=0,1,...)$, where $s=1.006$ for three identical bosons. Also, the ratio of two consecutive root-mean-square hyperradius, is exactly the square root of the energy ratios $e^{\pi/s}$.

\begin{figure}[htb!]
\centering
\includegraphics[width=10cm]{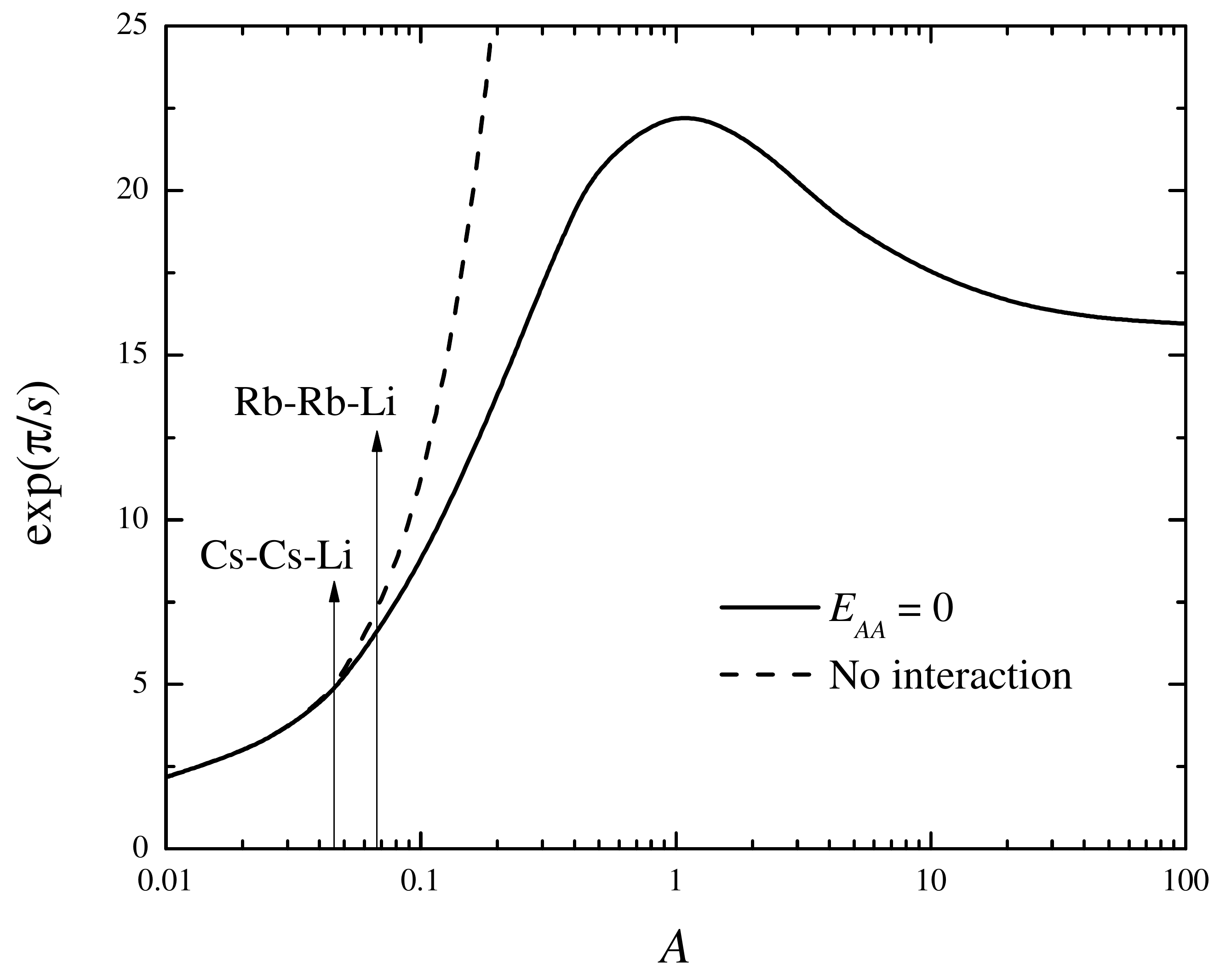}
\caption{Figure from Ref. \cite{YamPRA2013}. Scaling parameter $s$ as a function of the mass ratio $A=m_B/m_A$.}
\label{exps}
\end{figure}

Here, it is important to call the attention that the quantity $s$ depends on the mass ratio of the components of the system. For a system $AAB$, formed by two identical bosons of masses $m_A$ and a different particle with mass $m_B$, the change of $s$ with $A=m_B/m_A$ is showed in Fig. \ref{exps}, where it is plotted two distinct situations: the solid line shows the situation where the subsystem $AA$ has an energy equal to zero and the dashed line where there is no interaction between $AA$.
Figure (\ref{exps}) shows that for a borromean situation (we have a three-body bound states with all pairs unbound) the case of three identical masses, $A=1$, represents the largest gap between two energy levels. For this reason, mass asymmetric systems may present the most interesting situation for an experimental observation of Efimov states as the detection of two subsequent energy levels is facilitated.

The next figure shows the emergence of Efimov states from the two-body energy cut $\epsilon_3=\epsilon_2$. The lines with symbols are results for the ground (crosses), first (squares) and second (diamonds) excited states. The triangles and circles are virtual states. Virtual states are bound states in the second Riemann energy sheet. Every time we have at least one two-body susbsystem bound, the pole of the three-body bound state pass through the two-body energy cut defined by the two-body bound state and go to second Riemann sheet becoming a virtual state. The method used to calculate such states is beyond the scope of theses lectures, but a straightforward deduction can be found in this reference paper \cite{YamPRA2002}.
\begin{figure}[htb!]
\centering
\includegraphics[width=10cm]{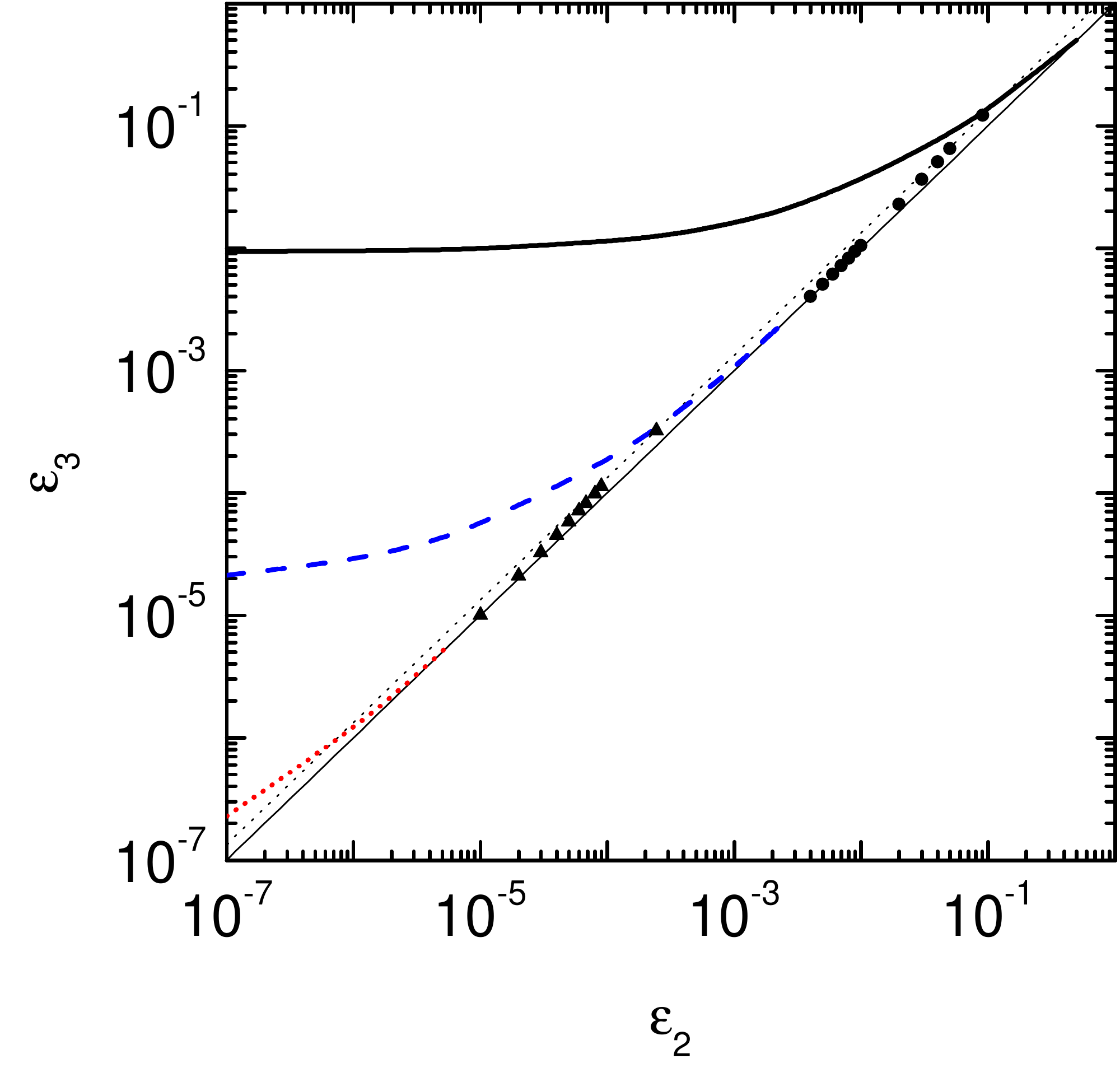}
\caption{Figure from Ref. \cite{YamPRA2002}. Efimov states emerging from the two-body energy cut as $\epsilon_2$ is decreased. The symbols are virtual states.}
\label{efimov}
\end{figure}

\subsection{Universality}

The use of zero-range potential raises a natural question about the consequences of changing the shape of the short-range potential. The answer to this question is quite surprising as the answer is: practically nothing will change. The independence of the observables on the type of the interaction potential is a property called {\it universality}. The physical reason for this universality is that the weakly-bound systems are spatially extended and, for this reason, dominated by the tail of the wave function\footnote{It is important to note that the potential should decrease faster than $\sim 1/r^2$. We may have very extended systems like, e.g., Rydberg atoms (excited atom with one or more electron with a very high principal quantum number) that do not enter in the definition of universality given here as, in this case, the Coulomb potential decays too slow as $r\to\infty$}. Thus, the details of the potential at short distances will not affect the observables. 

There is a large class of systems which we can include in the category of ``universal systems''. In these systems, a large part of the wave function extends through the classical forbidden region, in such a way the phenomena associated with universality are purely coming from quantum mechanics. We can find genuine universal systems in nature like halo nuclei or helium trimers \cite{universalsystems}, but we can artificially produce them inside ultracold traps using a Feshbach resonance.

\paragraph{Feshbach resonance} A Feshbach resonance is a mechanism which allows the tuning of the two-body scattering length. In the next lines we will show how this tuning becomes possible. We will follow the book by C. J. Pethick and H. Smith \cite{BEC}. Let us consider two atoms with nuclear spins $I_1$ and $I_2$ and electronic spin $s=1/2$. Then the number of hyperfine states (hs) is given by $(2s+1)(2I_1+1)(2s+1)(2I_2+1)=2(2I_1+1)(2I_2+1)$. Each hs will be called a {\it channel} and will be associated to a given greek letter. Then, in principle, two atoms in the state $|\alpha\beta\rangle$ may scatter to $|\alpha^\prime\beta^\prime\rangle$. This is a {\it multichannel} problem. In the absence of interaction between both atoms and in the presence of a magnetic field $\vec B$, we have:

\begin{equation}
H=H_0+H_{{\rm spin}}^{(1)}+H_{{\rm spin}}^{(2)}=\frac{p^2}{2\mu}+H_{{\rm spin}}^{(1)}+H_{{\rm spin}}^{(2)},
\end{equation}
where 
\begin{eqnarray}
\nonumber
&&H_{{\rm spin}}|\alpha\rangle=\epsilon_\alpha|\alpha\rangle\\
\nonumber
&&H|\alpha\beta\rangle=E_{\alpha\beta}|\alpha\beta\rangle;\,\,E_{\alpha\beta}=\frac{\hbar^2k_{\alpha\beta}^2}{2\mu}+\epsilon_\alpha+\epsilon_\beta.
\end{eqnarray}
As the total energy should be conserved we have that
\begin{eqnarray}
\nonumber
&&E_{\alpha^\prime\beta^\prime}+\epsilon_{\alpha^\prime}+\epsilon_{\beta^\prime}=\frac{\hbar^2k_{\alpha^\prime\beta^\prime}^{^\prime2}}{2\mu}+\epsilon_{\alpha^\prime}+\epsilon_{\beta^\prime}=\frac{\hbar^2k_{\alpha\beta}^2}{2\mu}+\epsilon_\alpha+\epsilon_\beta,\\ 
&&\frac{\hbar^2k_{\alpha^\prime\beta^\prime}^{^\prime2}}{2\mu}=\frac{\hbar^2k_{\alpha\beta}^2}{2\mu}+\epsilon_\alpha+\epsilon_\beta-(\epsilon_{\alpha^\prime}+\epsilon_{\beta^\prime}).
\label{energybalance}
\end{eqnarray}
Then if $k_{\alpha^\prime\beta^\prime}^{^\prime2}\leq0$ the channel is called closed. 

We may now define two subspaces. The subspace $P$ contains the open channels and the subspace $Q$ the closed channels. Two atoms start a collision in an open channel with a total energy $E=\frac{\hbar^2k_{\alpha\beta}^2}{2\mu}+\epsilon_\alpha+\epsilon_\beta$, given by the first three terms on the right-hand-side of Eq. (\ref{energybalance}). As this is an open channel, the total energy should be $E\geq E_{{\rm th}}$, where $E_{{\rm th}}\equiv \epsilon_\alpha+\epsilon_\beta$ is the threshold energy. Consider that a state vector may be written as:
\begin{equation}
|\Psi\rangle=|\Psi_P\rangle+|\Psi_Q\rangle,
\end{equation}
where $\hat{P}|\Psi\rangle=|\Psi_P\rangle$ and $\hat{Q}|\Psi\rangle=|\Psi_Q\rangle$ and the projection operators $\hat{P}$ and $\hat{Q}$ satisfy the following properties $\hat{P}+\hat{Q}=1$, $\hat{P}\hat{Q}=0$ and $\hat{Q}\hat{P}=0$. 

Multiplying the Schr\"odinger equation $H|\Psi\rangle=E|\Psi\rangle$ by $\hat{P}$ and $\hat{Q}$ we have that
\begin{eqnarray}
\label{sub1}
&&(E-H_{PP})|\Psi_P\rangle=H_{PQ}|\Psi_Q\rangle \\
&&(E-H_{QQ})|\Psi_Q\rangle=H_{QP}|\Psi_P\rangle,
\label{sub2}
\end{eqnarray}
where $H_{PP}=\hat{P}H\hat{P}$, $H_{QQ}=\hat{Q}H\hat{Q}$, $H_{PQ}=\hat{P}H\hat{Q}$ and $H_{PP}=\hat{Q}H\hat{P}$. The formal solution of Eq. (\ref{sub2}) for an outgoing wave is
\begin{equation}
|\Psi_Q\rangle=(E-H_{QQ}+i\delta)^{-1}H_{QP}|\Psi_P\rangle.
\label{psiqsol}
\end{equation}
Replacing Eq. (\ref{psiqsol}) in (\ref{sub1}), we have
\begin{equation}
(E-H_{PP})|\Psi_P\rangle=\left[H_{PQ}(E-H_{QQ}+i\delta)^{-1}H_{QP}\right]|\Psi_P\rangle,
\label{eqfesh}
\end{equation}
where the term inside the square brackets describe the Feshbach resonance. We will call it by 
\begin{equation}
H_{PP}^\prime\equiv H_{PQ}(E-H_{QQ}+i\delta)^{-1}H_{QP}.
\label{hpp}
\end{equation}
$H_{PP}^\prime$ shows the appearance of an effective interaction in the $P$ subspace due to transitions from $P$ to $Q$ and then back to $P$. This is a non-local potential in the open channel. The diagonal part of the Hamiltonian, $H_{PP}$, may be written as a sum of the free Hamiltonian, $H_0$ (which contains only the kinetic energy of the relative motion) and the other energies coming from the fine or hyperfine terms, $U_1$ (the interaction in $P$ space). Calling also $H_{PP}^\prime\equiv U_2$ and $U\equiv U_1+U_2$, we have from Eq. (\ref{eqfesh})
\begin{equation}
(E-H_0-U)|\Psi_P\rangle=0.
\end{equation}

\noindent\rule[0.5ex]{\linewidth}{1pt}
{\bf Exercise 13:} Verify Eqs. (\ref{sub1}) and (\ref{sub2}).

\noindent\rule[0.5ex]{\linewidth}{1pt}

It is clear that the coupling between the open and closed channels generates an effective interaction in the open channel. However, we still have to show how the two-body scattering length may be tuned with the magnetic field. To show this, let us start writing the $T$-operator for the interaction $U$ (Eq. (\ref{tmatrixg0})):
\begin{equation}
\nonumber
T=U+UG_0^{(+)}T,
\end{equation}
where the formal solution is 
\begin{equation}
T=(1-UG_0^{(+)})^{-1}U=U(1-G_0^{(+)}U)^{-1}
\label{formalT}
\end{equation}

\noindent\rule[0.5ex]{\linewidth}{1pt}
{\bf Exercise 14:} Verify the second equality of Eq. (\ref{formalT}). Note that $U$ does not commute with $(1-UG_0^{(+)})^{-1}$, but $UG_0^{(+)}$ does.

\noindent\rule[0.5ex]{\linewidth}{1pt}

Writing the free Green function explicitly in Eq. (\ref{formalT}) we can write
\begin{eqnarray}
\nonumber
&&T=\left(1-U(E-H_0+i\delta)^{-1}\right)^{-1}U=\left\{\left[(E-H_0+i\delta)-U\right]\left(E-H_0+i\delta\right)^{-1}\right\}^{-1}U\\
&&T=(E-H_0+i\delta)(E-H_0+i\delta-U)^{-1}U.
\label{id1}
\end{eqnarray}
Calling $A=E+i\delta-H_0-U_1$ and $B=U_2$ and using the identity $(A-B)^{-1}=A^{-1}(1+(B(A-B)^{-1})$ we can write the following equation:
\begin{equation}
(E+i\delta-H_0-U)^{-1}=(E+i\delta-H_0-U_1)^{-1}\left[1+U_2(E+i\delta-H_0-U)^{-1}\right].
\label{id2}
\end{equation}
Then, replacing Eq. (\ref{id2}) in (\ref{id1}) we can write
\begin{equation}
T=T_1+(1-U_1G_0^{(+)})^{-1}U_2(1-G_0^{(+)}U)^{-1},
\label{id3}
\end{equation}
where $T_1=U_1+U_1G_0^{(+)}T_1$.

\noindent\rule[0.5ex]{\linewidth}{1pt}
{\bf Exercise 15:} Verify Eq. (\ref{id3}).

\noindent\rule[0.5ex]{\linewidth}{1pt}

Let us now calculate the following matrix elements between plane wave states with relative momenta $\vec{k}$ and $\vec{k}\,^\prime$:
\begin{equation}
\langle\vec{k}\,^\prime|T|\vec{k}\rangle=\langle\vec{k}\,^\prime|T_1|\vec{k}\rangle+\langle\vec{k}\,^\prime|(1-U_1G_0^{(+)})^{-1}U_2(1-G_0^{(+)}U)^{-1}|\vec{k}\rangle.
\end{equation}

From the Lippmann-Schwinger equation, Eq. (\ref{LS}), we see that the operator $(1-G_0^{(+)}U)^{-1}$ applied to the plane wave, $|\vec{k}\rangle$, returns the scattered wave, $|\Psi_{\vec{k}}^{(+)}\rangle$:
\begin{eqnarray}
\nonumber
&&|\Psi_{\vec{k}}^{(+),U}\rangle=|\vec{k}\rangle+G_0^{(+)}U|\Psi_{\vec{k}}^{(+)}\rangle\\
&&|\Psi_{\vec{k}}^{(+),U}\rangle=(1-G_0^{(+)}U)^{-1}|\vec{k}\rangle,\\ \nonumber \\ 
&&\langle\vec{k}\,^\prime|(1-U_1G_0^{(+)})^{-1}=\left[(1-U_1G_0^{(-)})^{-1}|\vec{k}\,^\prime\rangle\right]^\dagger=\left[|\Psi_{\vec{k}\,^\prime}^{(-),U_1}\rangle\right]^\dagger,
\end{eqnarray}
where $G_0^{(-)}=(E-H_0-i\delta)^{-1}$. Thus,
\begin{equation}
\langle\vec{k}\,^\prime|T|\vec{k}\rangle=\langle\vec{k}\,^\prime|T_1|\vec{k}\rangle+\langle\Psi_{\vec{k}\,^\prime}^{(-),U_1}|U_2|\Psi_{\vec{k}}^{(+),U}\rangle.
\label{generalscat}
\end{equation}

Eq. (\ref{generalscat}) is the general equation for the scattering amplitude in the $P$ space. The term involving $T_1$ is the matrix in $P$ subspace if the transitions to the $Q$ subspace are neglected. Let us now simplify a little our calculations. Consider that the relative velocity of the incoming particles is extremely low $k\approx0$. In this situation there are no phase factors in such a way we may neglect the difference between scattering states with incoming and outgoing spherical waves. Then:
\begin{equation}
|\Psi_{\vec{k}\,^\prime}^{(-),U_1}\rangle=|\Psi_{\vec{k}}^{(+),U}\rangle\equiv|\Psi_0\rangle.
\end{equation}
In this limit, the second matrix element on the right-hand side reads:
\begin{eqnarray}
\nonumber
&&\langle\Psi_0|H_{PQ}(E-H_{QQ}+i\delta)^{-1}H_{QP}|\Psi_0\rangle=\sum_n\langle\Psi_0|H_{PQ}(E-H_{QQ}+i\delta)^{-1}|\Psi_n\rangle\langle\Psi_n|H_{QP}|\Psi_0\rangle\\ 
&&\sum_n\frac{\langle\Psi_0|H_{PQ}|\Psi_n\rangle\langle\Psi_n|H_{QP}|\Psi_0\rangle}{E-E_n}=\sum_n\frac{\left|\langle\Psi_n|H_{QP}|\Psi_0\rangle\right|^2}{E-E_n}.
\end{eqnarray}
Using that $T(\vec{k}\,^\prime=0,\vec{k}=0)=\frac{4\pi\hbar^2}{m}a$, where $m$ is the reduced mass and $a$ the two-body scattering length, we have that:
\begin{equation}
\frac{4\pi\hbar^2}{m}a=\frac{4\pi\hbar^2}{m}a_{{\rm bg}}+\frac{\left|\langle\Psi_{{\rm RES}}|H_{QP}|\Psi_0\rangle\right|^2}{E-E_{{\rm RES}}}\,\,\,(E\sim E_{{\rm RES}}).
\label{fesh1}
\end{equation}
Eq. (\ref{fesh1}) was written considering that $E$ is close to a given energy $E_n\equiv E_{{\rm RES}}$ in such a way we can  disregard the other terms of the sum. The scattering length $a_{{\rm bg}}$ is a non-resonant scattering length that results from the background collision in the open channel $P$. 

Considering the relative velocity approximately zero, the energy in the entrance channel is given by the energies of the hyperfine states $E=\epsilon_\alpha+\epsilon_\beta$. All energies here depend on the magnetic field, $B$ and let us consider that the denominator $E-E_{{\rm RES}}$ vanishes for a given $B=B_0$. Expanding this term we have:
\begin{equation}
E-E_{{\rm RES}}=\left[\frac{\partial\epsilon_\alpha}{\partial B}+\frac{\partial\epsilon_\beta}{\partial B}-\frac{\partial E_{\rm RES}}{\partial B}\right](B-B_0),
\end{equation}
the magnetic moment is defined as $\mu=-\frac{\partial\epsilon}{\partial B}$, thus
\begin{equation}
E-E_{{\rm RES}}=(\mu_{{\rm RES}}-\mu_\alpha-\mu_\beta)(B-B_0).
\label{denom}
\end{equation}
Replacing Eq. (\ref{denom}) in Eq. (\ref{fesh1}) we have the form of the Feshbach resonance showed in many papers:
\begin{eqnarray}
\nonumber
&&a=a_{{\rm bg}}+\frac{m}{4\pi\hbar^2}\frac{\left|\langle\Psi_{{\rm RES}}|H_{QP}|\Psi_0\rangle\right|^2}{(\mu_{{\rm RES}}-\mu_\alpha-\mu_\beta)}\frac{1}{(B-B_0)}\\ \nonumber
&&a=a_{{\rm bg}}\left(1+\frac{\Delta B}{(B-B_0)}\right);\,\,\,\,\Delta B\equiv \frac{m}{4\pi\hbar^2}\frac{\left|\langle\Psi_{{\rm RES}}|H_{QP}|\Psi_0\rangle\right|^2}{(\mu_{{\rm RES}}-\mu_\alpha-\mu_\beta)}.
\label{feshbach}
\end{eqnarray}
The quantity $\Delta B$ is called width parameter. We can see from Eq. (\ref{feshbach}) that tuning the magnetic field the two-body scattering length can be made positive or negative passing through the infinite. This allows an incredible freedom to study the correlations between the scales. The experimental realization of the Feshbach resonance opened new horizons to the few-body area.

Returning to the discussion of universality, as the observables do not depend on the details of the short-range potential all results may be described only by the observables themselves. Suppose a three-dimensional system formed by three-identical bosons forming a bound state with energy $E_3$, and with two-body subsystems with energies $E_2$. If this system satisfy the condition of universality, any observable can be described by a universal function $\cal{F}$ as:
\begin{equation}
{\cal O}(E,E_2,E_3)=(E_3)^\eta{\cal F}(E,E_2,E_3),
\label{unifunc}
\end{equation}
where ${\cal O}(E,E_2,E_3)$ is any three-body observable at a given energy $E$ with dimension of $(E_3)^\eta$. Arbitrarily we chose as two- and three-body scales the binding energies $E_2$ and $E_3$. The necessity to add a three-body scale is directly related to the possibility of collapse of the three-body system. Let us clarify what we understand by collapse. Consider that the three-body system interacts only by a two-body interaction.  Then, decreasing the range of the potential, but keeping the two-body energy fixed, the three-body binding energy can be made arbitrarily large, i.e., the system is unbound from below - this is the collapse. This problem was firstly studied by L. H. Thomas \cite{thomas} and today it is known by Thomas collapse.

The Thomas collapse is also related to the need of inclusion of a new scale to describe the problem. An infinite large three-body binding energy is not a comfortable situation. In order to solve this divergence we have to insert a cutoff in our theory or, equivalently, a new scale as showed in Eq. (\ref{eqsub32a}). This new scale was arbitrarily chosen as being $E_3$ in Eq. (\ref{unifunc}), but it could be another three-body observable. If the collapse of three-body scale does not exists, then any observable of the system may be described by a universal function dependent only on the two-body scale. This absence of collapse happens, for example, in a three-identical fermions interacting in $s$-wave: due to the Pauli principle the collapse is avoided. 
\begin{figure}[htb!]
\centering
\includegraphics[width=10cm]{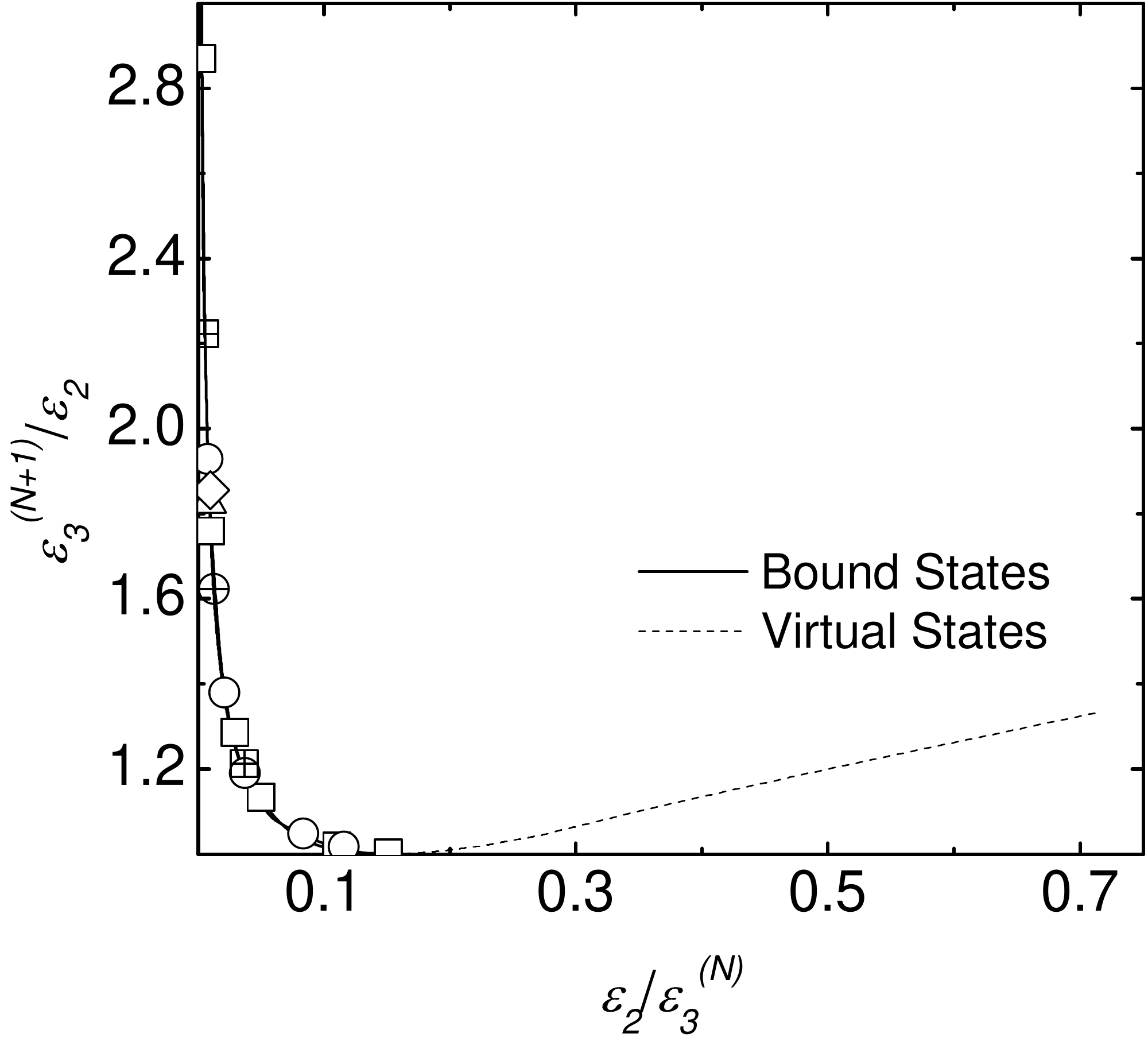}
\caption{Universal scaling plot for the Efimov spectrum. Figure from Ref. \cite{YamPRA2002}. The several points inside the plot are calculations with realistic potentials - note that we have a very good agreement with the zero-range potential. The dashed line is a bound state on the second Riemmann sheet, they are called virtual states (virtual states are beyond the scope of these lectures).}
\label{scalingplot}
\end{figure}

Let us return to Fig. \ref{efimov} where we plotted the Efimov states emerging from the two-body energy cut. According to Eq. (\ref{unifunc}), any excited three-body energy $\epsilon_3^{(N+1)}$ might be described by a universal function that depends uniquely on a two and a three-body scales. We can choose these scales as being the two-body energy, $\epsilon_2$, and the previous three-body energy, $\epsilon_3^{(N)}$. This scaling function is plotted in Fig. \ref{scalingplot}.

It is worth to mention, that practically there is no difference if we use $N=0,1,2...$ to construct the figure. The largest difference happens when considering the ground state ($N=0$). The reason is that this state is smaller than the excited ones and the condition for the universality is worse satisfied. The point where the bound states disappear inside the two-body energy cut is also universal and it is given by $\epsilon_2/\epsilon_3^{(N)}=0.145$.

\section{Final Remarks}

These lectures are primarily dedicated to the students. I tried to include the concepts which I judged to be important to start a study in momentum space techniques in Few-Body Physics. Excepting the last part, which is just an overview of some works of our Brazilian group, the other sections are very detailed and I tried to be complete. More information about our papers may be found at www.ift.unesp.br/users/yamashita (ugly but useful). Suggestions are welcome. Please write your feedback to yamashita@ift.unesp.br.

\end{document}